\documentclass[3p,12pt]{elsarticle}
\usepackage{hyperref}
\usepackage{amssymb}
\usepackage{amsmath}
\usepackage{algorithm}
\usepackage{algpseudocode}
\usepackage{enumitem}
\usepackage{cleveref}
\usepackage{setspace}
\usepackage{booktabs}
\usepackage{subfigure}
\usepackage{xcolor}
\usepackage{multirow}
\usepackage[T1]{fontenc}
\journal{Engineering Structures}

\begin{document}
%\begin{linenumbers}
\begin{frontmatter}

%\linenumbers


\title{Reduced-Order Physics-Informed Neural Network with Adaptive Basis Refinement for Structural Identification} %% Article title


\author[mymainaddress]{Rui Zhang\corref{mycorrespondingauthor}}
\cortext[mycorrespondingauthor]{Corresponding author}
\ead{rui.zhang@ibk.baug.ethz.ch}
		
\author[mymainaddress]{Konstantinos Vlachas}
		
\author[mymainaddress]{Eleni Chatzi}
	
\affiliation[mymainaddress]{
    organization={Department of Civil, Environmental and Geomatic Engineering, ETH Zürich},
    addressline={Stefano-Franscini-Platz 5},
    city={Zürich},
    postcode={8093},
    country={Switzerland}
    }

%% Abstract
\begin{abstract}
Physics-informed neural networks (PINNs) provide a flexible framework for solving forward and inverse problems. However, their direct application to structural dynamics remains limited by high system dimensionality and by model-form errors arising from incomplete physics. Reduced-order models (ROMs) can alleviate the dimensionality bottleneck, yet existing PINN and ROM couplings typically rely on fixed reduced subspaces, target forward simulations, or assume complete physics, which restricts their use for inverse identification under parametric variability or incomplete system knowledge. To address these limitations, this work proposes a Reduced-Order Physics-Informed Neural Network (RO-PINN) framework with adaptive basis refinement for structural identification under known and incomplete physics. Via projection, the reduced governing equations are embedded directly into the PINN loss, facilitating learning in a low-dimensional latent space. Simultaneously, an adaptive scheme updates the projection basis during training so that the latent space is progressively realigned with evolving structural parameters or learned residual restoring forces. This progressive realignment reduces basis-mismatch errors and limits their influence on the inferred residual force. The proposed method is validated on a four-story steel frame with nonlinear hysteretic braces under sparse and noisy measurements. Results show parameter identification comparable to or more accurate than Bayesian model updating with lower computational cost in the considered cases, recovery of unmodeled nonlinear restoring forces under incomplete physics, and joint identification of residual restoring forces and structural parameters within the same framework. Overall, the proposed RO-PINN provides a unified framework for structural identification by integrating reduced-order modeling, adaptive basis refinement, and physics-informed learning within a single formulation.
\end{abstract}

%% Keywords

\end{frontmatter}
\section{Introduction}
\label{s1}

Accurate and efficient system identification in structural dynamics is of central importance across the physical and engineering sciences, underpinning applications such as structural health monitoring, digital twins, and resilient infrastructure design~\cite{sun2024approach,hou2021review}.

Classical approaches are rooted in physics-based modeling, most notably finite-element (FE) analysis and model updating~\cite{chatzi2016identification,zarate2008finite}. These methods infer uncertain parameters from vibration measurements by repeatedly solving the underlying high-dimensional FE model, which can make the inverse problem computationally demanding~\cite{lye2021sampling,behmanesh2015hierarchical}. More fundamentally, conventional parameter updating is limited when the assumed governing model is incomplete. If relevant physical contributions are absent from the prescribed model class, adjusting existing parameters alone cannot explicitly represent the missing physics.

The recent emergence of scientific machine learning (SciML) has offered new pathways for integrating physics-based formulations with data-driven models. A prominent example is Physics-Informed Neural Networks (PINNs)~\cite{raissi2019physics,karniadakis2021physics}, which incorporate governing equations and available observations into neural-network training. PINNs provide a flexible framework for forward and inverse problems~\cite{zhang2024dual,wang2024structural,zhang2020physics,mishra2022estimates} and may also be extended to infer unknown physical contributions directly from data~\cite{reyes2021learning,liu2025physics}. However, their application to large-scale nonlinear structural systems remains challenging. Training cost increases with the dimensionality of the physical state, while stiff and multi-scale dynamics may lead to slow or difficult optimization~\cite{wang2021eigenvector,wang2021understanding}. Performance can also be sensitive to the relative weighting of data and physics losses~\cite{wang2022and,zhang2024physics}.

To reduce the computational burden of physics-informed learning, this work leverages projection-based Reduced-Order Modeling (ROM)~\cite{benner2017model}. ROMs approximate high-dimensional dynamics in low-dimensional subspaces while retaining the dominant response features of the underlying system~\cite{azam2013investigation, patsialis2020reduced}. Classical techniques such as Proper Orthogonal Decomposition (POD)~\cite{vlachas2021local}, Proper Generalized Decomposition~\cite{chinesta2020proper}, and Component Mode Synthesis (CMS)~\cite{allen2020substructuring} have been widely used to reduce the computational cost of high-fidelity structural and multiphysics models. Embedding projection-based reduced equations within a PINN offers a natural means of reducing the dimensionality of the learning problem while preserving an explicit connection to the governing dynamics.

Recent studies have explored combinations of reduced-order models and physics-informed learning. Projection-based approaches have incorporated POD--Galerkin reduced equations into PINN formulations for parameter identification~\cite{hijazi2023pod}, while other methods reduce the governing equations or operate in latent representations obtained through autoencoders~\cite{pan2024ro,kim2022fast,chen2021physics}. Related applications include fluid dynamics, heat conduction, and low-dimensional dynamical systems, where physics-informed networks enforce reduced dynamics or learn reduced coordinates~\cite{luong2024novel,hong2024physics,sibuet2025discrete, yin2025physics}. Nevertheless, most existing reduced-order physics-informed formulations rely on reduced representations constructed before training and assume that the prescribed governing model remains valid throughout the inverse solution. For structural identification, these assumptions can become restrictive when the inferred parameters move away from the nominal configuration or when relevant restoring-force contributions are missing from the assumed model.

To address these issues, we propose a Reduced-Order Physics-Informed Neural Network (RO-PINN) with adaptive basis refinement for inverse structural dynamics. The proposed framework embeds projected governing equations directly into the physics-informed loss so that the inverse problem is solved in a low-dimensional representation. Rather than retaining a fixed reduced basis, the reduced subspace is updated as the inferred physical quantities evolve, thereby reducing basis-mismatch effects during identification. The formulation accommodates parameter identification under known physics, residual-force identification under incomplete physics, and their simultaneous estimation within a unified reduced-order framework.

The key contributions of this work are summarized as follows:
\begin{enumerate}[leftmargin=*]

    \item An adaptive reduced-basis refinement strategy coupled with the inverse optimization, in which the reduced subspace and projected operators are updated as the inferred physical quantities evolve, thereby reducing basis-mismatch effects during identification.

    \item A unified reduced-order physics-informed formulation that accommodates parameter identification under known physics, residual-force identification under incomplete physics, and their simultaneous estimation within the same inverse-learning framework.

    \item A reduced nonlinear-dynamics formulation that operates on low-dimensional structural coordinates and can additionally incorporate
    reduced internal variables governed by known evolution laws, enabling the treatment of high-dimensional nonlinear structural systems with hysteretic behavior.

\end{enumerate}

The remainder of the paper is organized as follows. Section~\ref{s2} formulates the structural inverse problem and motivates the reduced-order physics-informed framework. Section~\ref{s3} presents the RO-PINN methodology, including the governing formulation, loss construction, adaptive basis refinement, and optimization procedure. Section~\ref{s4} presents the numerical studies, and Section~\ref{s5} summarizes the main findings, limitations, and future directions.

%%%%%%%%%%%%%%%%%%%%%%%%%%%%%%%%%%%%%%%%%%%%%%%%%%%%%%%%%%%%%%%%%%%%%%
\section{Problem Statement and Motivation}
\label{s2}

\subsection{Governing Equations and Problem Statement}
\label{s2.1}

A general structural dynamical system is considered, characterized by a parameter vector $\boldsymbol{\gamma}\in\mathbb{R}^{N_p}$ containing the structural parameters of interest. Its dynamic response is governed by
\begin{equation}\label{GE_u}
\mathbf{M}\ddot{\mathbf{u}}(t)
+
\mathbf{g}
\left(
\mathbf{u}(t),
\dot{\mathbf{u}}(t),
\boldsymbol{\gamma}
\right)
=
\mathbf{P}(t),
\end{equation}
where $\mathbf{u}(t)\in\mathbb{R}^{n}$ is the displacement vector, $\mathbf{M}\in\mathbb{R}^{n\times n}$ is the mass matrix, and $\mathbf{P}(t)\in\mathbb{R}^{n}$ denotes the external excitation. The restoring force $\mathbf{g}$ includes stiffness and damping contributions and may account for material or geometric nonlinearities. The quantity $n$ denotes the full-order dimension of the spatially discretized system.

In practice, the assumed governing equations may not provide a complete representation of the physical system. The restoring force is therefore decomposed into a known-form contribution and an unknown residual contribution:
\begin{equation}\label{GE_u_augmented}
\mathbf{M}\ddot{\mathbf{u}}(t)
+
\mathbf{g}^{\ast}
\left(
\mathbf{u}(t),
\dot{\mathbf{u}}(t);
\boldsymbol{\gamma}
\right)
+
\mathbf{R}(t)
=
\mathbf{P}(t),
\end{equation}
where $\mathbf{g}^{\ast}(\cdot;\boldsymbol{\gamma})$ denotes the known or assumed restoring-force model parameterized by $\boldsymbol{\gamma}$, and $\mathbf{R}(t)$ denotes an unknown residual restoring force associated with missing or unmodeled physics.

Equation~\eqref{GE_u_augmented} distinguishes uncertainty in the parameters of the known-form model from model-form error represented by the residual force. These two classes of unknowns give rise to the three inverse-problem configurations considered in this work:
\begin{enumerate}[leftmargin=*]

    \item Parameter identification with known physics:
    the functional form of the restoring force is assumed known, but some structural or material parameters are unknown. In this case,
    $\mathbf{R}(t)\equiv\mathbf{0}$ and the objective is to identify $\boldsymbol{\gamma}$.

    \item Residual-force identification under incomplete physics: part of the restoring force is not represented by the assumed model. The objective is therefore to identify the residual restoring-force contribution $\mathbf{R}(t)$ associated with the missing physics.

    \item Joint force--parameter identification: both classes of unknowns are present, and $\boldsymbol{\gamma}$ and $\mathbf{R}(t)$ are inferred simultaneously.

\end{enumerate}

In all three configurations, the objective is to identify the unknown quantities and reconstruct the structural response from available measurements while maintaining consistency with the governing dynamics.

\subsection{Motivation for Reduced-Order Physics-Informed Learning}
\label{s2.2}

Model-form error presents a major challenge for conventional modeling and identification methods~\cite{kamariotis2025consistent}. Finite-element model updating and projection-based reduced-order modeling generally rely on an assumed form of the governing equations and primarily adjust parameters already contained in that model~\cite{hollins2026projection,vlachas2025utility}. Physics-informed neural networks (PINNs), in contrast, provide a flexible means of inferring unknown physical contributions directly from data and governing equations, but their computational cost can become substantial for high-dimensional structural systems.

These considerations motivate a framework that combines the flexibility of PINNs with the computational efficiency of reduced-order modeling. The proposed RO-PINN embeds reduced governing equations within the physics-informed loss so that inverse identification is performed in a low-dimensional representation. Because the reduced basis constructed from an initial or nominal model may become inconsistent as the inferred parameters and/or residual forces evolve, adaptive basis refinement is introduced to progressively realign the reduced subspace with the identified system. The framework therefore accommodates parameter identification, residual-force identification, and their joint estimation within a unified reduced-order physics-informed formulation. The methodological details are presented in Section~\ref{s3}.

%%%%%%%%%%%%%%%%%%%%%%%%%%%%%%%%%%%%%%%%%%%%%%%%%%%%%%%%%%%%%%%%%%%%%%
\section{RO-PINN Methodology: Reduced-Order Physics-Informed Neural Networks}
\label{s3}

This section presents the proposed Reduced-Order Physics-Informed Neural Network (RO-PINN) framework. Section~\ref{s3.1} introduces the overall methodology and its key components. Section~\ref{s3.2} formulates the physics-informed loss function by combining measurement data and reduced-order dynamics. Section~\ref{s3.3} describes the adaptive basis-refinement strategy used to maintain reduced-order accuracy and physics consistency. Finally, Section~\ref{s3.4} presents the optimization procedure and reconstruction of the full-field structural response.

\subsection{Overview of the Framework}
\label{s3.1}

\begin{figure}[h!]
    \centering
    \includegraphics[scale=0.32]{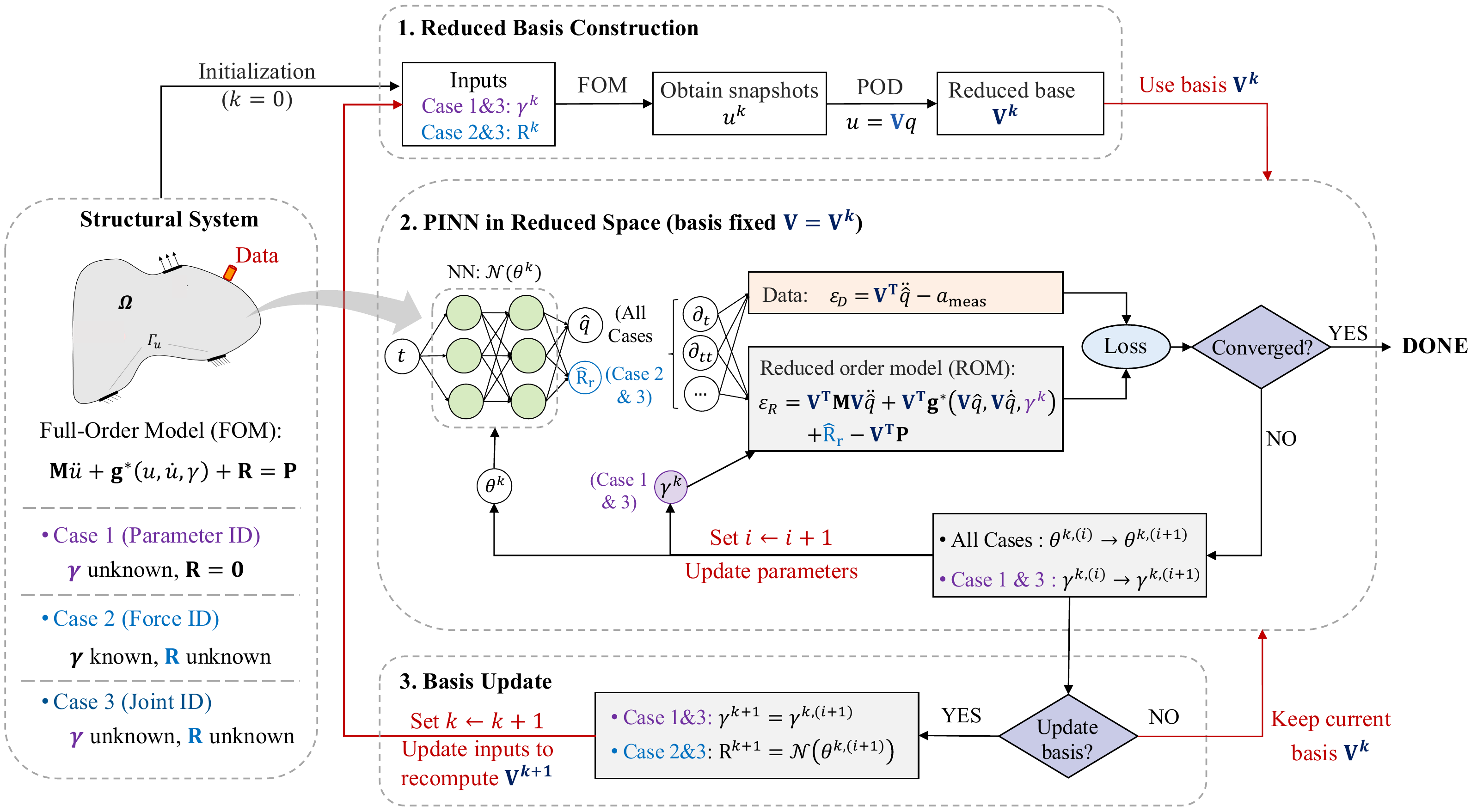}
\caption{Overview of the proposed RO-PINN framework with adaptive basis refinement. At each outer iteration $k$, the inverse problem is solved using the current basis $\mathbf{V}^{k}$. With $\mathbf{V}^{k}$ fixed, the PINN predicts the reduced coordinates $\hat{\mathbf{q}}(t)$ and, when residual-force identification is considered, an additional residual-force representation. The loss combines measurement mismatch and reduced-order equilibrium residuals to update the neural-network parameters $\boldsymbol{\theta}$ and, when applicable, the structural parameters $\boldsymbol{\gamma}$. The resulting inverse estimates are then introduced into the full-order model (FOM) to generate response snapshots and construct the updated basis $\mathbf{V}^{k+1}$ through POD. Here, $\mathbf{g}^{\ast}(\cdot;\boldsymbol{\gamma})$ denotes the known-form restoring force and $\mathbf{R}(t)$ denotes the residual restoring-force contribution associated with missing physics.
}
    \label{fig:ROM_PIPNNs}
\end{figure}

The proposed RO-PINN framework, illustrated in Figure~\ref{fig:ROM_PIPNNs}, integrates physics-informed neural networks (PINNs), reduced-order modeling (ROM), and adaptive basis refinement within a unified inverse-analysis framework. The inverse settings introduced in Section~\ref{s2.1}, including parameter identification under known physics, residual-force identification under incomplete physics, and their joint estimation, are treated within the same reduced-order learning procedure.

During the first outer iteration, $k=0$, a reduced basis $\mathbf{V}^{0}\in\mathbb{R}^{n\times r}$ is constructed from a nominal, possibly incomplete, model. More generally, for a given basis $\mathbf{V}^{k}$, the inverse problem is first solved to obtain the current estimates of the unknown quantities. These estimates are then introduced into the full-order model in Eq.~\eqref{eq:basis_construction_k}, whose response is used to generate snapshots for constructing the updated basis $\mathbf{V}^{k+1}$:
\begin{equation}\label{eq:basis_construction_k}
\mathbf{M}\ddot{\mathbf{u}}^{k}(t)
+
\mathbf{g}^{\ast}\!\left(
\mathbf{u}^{k}(t),
\dot{\mathbf{u}}^{k}(t);
\boldsymbol{\gamma}^{k}
\right)
+
\mathbf{R}^{k}(t)
=
\mathbf{P}(t).
\end{equation}

The resulting full-order response is decomposed as
\begin{equation}\label{eq:proj_error}
\mathbf{u}^{k}(t)
=
\mathbf{V}^{k}\mathbf{q}^{k}(t)
+
\boldsymbol{\varepsilon}^{k}(t),
\end{equation}
where $\mathbf{q}^{k}(t)$ contains the response components resolved by the current basis and $\boldsymbol{\varepsilon}^{k}(t)$ denotes the unresolved component associated with the reduced-subspace approximation. Projecting Eq.~\eqref{eq:basis_construction_k} onto the current basis
$\mathbf{V}^{k}$ and substituting Eq.~\eqref{eq:proj_error} gives
\begin{equation}\label{eq:reduced_eq}
\mathbf{M}_r^{k}\ddot{\mathbf{q}}^{k}(t)
+
\mathbf{g}_r^{\ast,k}\!\left(
\mathbf{V}^{k}\mathbf{q}^{k}(t),
\mathbf{V}^{k}\dot{\mathbf{q}}^{k}(t);
\boldsymbol{\gamma}^{k}
\right)
+
{\varepsilon}_r^{k}(t)
+
\mathbf{R}_r^{k}(t)
=
\mathbf{P}_r^{k}(t),
\end{equation}
where
\[
\mathbf{M}_r^{k}
=
(\mathbf{V}^{k})^\top\mathbf{M}\mathbf{V}^{k},
\qquad
\mathbf{g}_r^{\ast,k}(\cdot)
=
(\mathbf{V}^{k})^\top\mathbf{g}^{\ast}(\cdot),
\]
\[
\mathbf{R}_r^{k}
=
(\mathbf{V}^{k})^\top\mathbf{R}^{k},
\qquad
\mathbf{P}_r^{k}
=
(\mathbf{V}^{k})^\top\mathbf{P}.
\]

The term ${\varepsilon}_r^{k}(t)$ collects the reduced consistency error associated with the unresolved response component $\boldsymbol{\varepsilon}^{k}(t)$:
\begin{equation}\label{eq:reduced_consistency_error}
{\varepsilon}_r^{k}(t)
=
(\mathbf{V}^{k})^\top
\mathbf{M}
\ddot{\boldsymbol{\varepsilon}}^{k}(t)
+
(\mathbf{V}^{k})^\top
\Big[
\mathbf{g}^{\ast}
\!\left(
\mathbf{V}^{k}\mathbf{q}^{k}
+
\boldsymbol{\varepsilon}^{k},
\mathbf{V}^{k}\dot{\mathbf{q}}^{k}
+
\dot{\boldsymbol{\varepsilon}}^{k};
\boldsymbol{\gamma}^{k}
\right)
-
\mathbf{g}^{\ast}
\!\left(
\mathbf{V}^{k}\mathbf{q}^{k},
\mathbf{V}^{k}\dot{\mathbf{q}}^{k};
\boldsymbol{\gamma}^{k}
\right)
\Big].
\end{equation}

For parameter identification, $\mathbf{R}^{k}(t)\equiv\mathbf{0}$ and $\boldsymbol{\gamma}^{k}$ is updated during training. For residual-force identification, the structural parameters are fixed and the residual force is predicted by the neural network. Depending on the number of unknown residual-force components, the network predicts either a reduced residual $\hat{\mathbf{R}}_r(t)$ or selected full-order residual-force components $\hat{\mathbf{R}}(t)$. When a reduced representation is used, the corresponding full-order residual-force representation is recovered before snapshot generation. In the joint setting, both $\boldsymbol{\gamma}^{k}$ and the residual-force estimate are updated simultaneously.

The neural network takes time $t$ as input and always predicts the reduced coordinates $\hat{\mathbf{q}}(t)$. Structural parameters $\boldsymbol{\gamma}$, when identified, are treated as additional trainable variables rather than network outputs. The full-field structural response is reconstructed as $\hat{\mathbf{u}}(t)=\mathbf{V}^{k}\hat{\mathbf{q}}(t)$. As the inverse estimates evolve, the current basis may become inconsistent with the identified system, increasing ${\varepsilon}_r^{k}$ and degrading the reduced-order representation. The adaptive basis-refinement strategy described in Section~\ref{s3.3} is therefore used to progressively realign the reduced subspace with the evolving system.

\subsection{Physics-Informed Loss Function with Reduced-Order Dynamics}
\label{s3.2}

For the inverse problems considered in this work, the RO-PINN is trained by minimizing a composite loss that combines the reduced-order governing equations with the available measurement data:
\begin{equation}
\label{eq:total_loss}
\mathcal{L}
=
\beta_{GE}\mathcal{L}_{GE}
+
\beta_{S}\mathcal{L}_{S},
\end{equation}
where $\mathcal{L}_{GE}$ enforces the reduced structural equilibrium and $\mathcal{L}_{S}$ measures the mismatch between predicted and measured structural responses. The loss weights $\beta_{GE}$ and $\beta_{S}$ may be prescribed manually or adapted automatically, for example using an NTK-based self-adaptive weighting strategy~\cite{zhang2024physics}.

For a fixed basis $\mathbf{V}^{k}$, the governing-equation loss is defined as the mean-squared reduced consistency residual introduced in Eq.~\eqref{eq:reduced_eq}:
\begin{equation}
\label{eq:loss_GE}
\begin{aligned}
\mathcal{L}_{GE}
=
\frac{1}{N_f}
\sum_{j=1}^{N_f}
\left\|
{\varepsilon}_{r}^{k}(t_j)
\right\|^2
=
\frac{1}{N_f}
\sum_{j=1}^{N_f}
\Bigg\|
\mathbf{P}_{r}^{k}(t_j)
-
\mathbf{M}_{r}^{k}\ddot{{\mathbf{q}}}(t_j)
-
\mathbf{g}_{r}^{\ast,k}
\!\left(
\mathbf{V}^{k}{\mathbf{q}}(t_j),
\mathbf{V}^{k}\dot{{\mathbf{q}}}(t_j);
\boldsymbol{\gamma}
\right)
-
{\mathbf{R}}_{r}^{k}(t_j)
\Bigg\|^2 ,
\end{aligned}
\end{equation}
where $\{t_j\}_{j=1}^{N_f}$ are temporal collocation points, and $\dot{{\mathbf{q}}}$ and $\ddot{{\mathbf{q}}}$ are obtained through automatic differentiation with respect to time. During training, the unknown exact quantities in
Eq.~\eqref{eq:loss_GE} are replaced by their current neural-network and
parameter estimates to evaluate the governing-equation residual.

For parameter identification under known physics, $\hat{\mathbf{R}}_{r}^{k}\equiv\mathbf{0}$ and the structural parameters $\boldsymbol{\gamma}$ are optimized as trainable variables. For residual-force identification, $\boldsymbol{\gamma}$ is prescribed and the residual force is predicted by the neural network. When many residual-force components are unknown, the network may predict $\hat{\mathbf{R}}_{r}$ directly in the reduced space; when only a limited number of components are unknown, the corresponding full-order force components $\hat{\mathbf{R}}$ may be predicted directly and projected as $\hat{\mathbf{R}}_{r}^{k}(t)=(\mathbf{V}^{k})^{\top}\hat{\mathbf{R}}(t)$. In the joint setting, $\boldsymbol{\gamma}$ and the residual force are inferred simultaneously. The joint problem considered here assumes that the two classes of unknowns act on different structural regions, reducing direct compensation between them and improving practical identifiability.

For systems whose nonlinear behavior follows known evolution laws, such as plasticity, hysteresis, viscoelasticity, or damage, the restoring force may depend on internal variables $\mathbf{z}(t)$ governed by
\begin{equation}
\dot{\mathbf{z}}
=
\mathbf{h}
\left(
\mathbf{u},
\dot{\mathbf{u}},
\mathbf{z};
\boldsymbol{\gamma}
\right).
\end{equation}
When such laws are available, the internal variables can also be represented in a reduced space as
$\mathbf{z}(t)\approx\mathbf{V}_{z}\mathbf{q}_{z}(t)$, and the network additionally predicts $\hat{\mathbf{q}}_{z}(t)$. Consistency with the prescribed evolution law is enforced through the additional governing-equation loss
\begin{equation}
\label{eq:loss_GE_z}
\mathcal{L}_{GE}^{(z)}
=
\frac{1}{N_f}
\sum_{j=1}^{N_f}
\left\|
\mathbf{V}_{z}\dot{{\mathbf{q}}}_{z}(t_j)
-
\mathbf{h}
\left(
\mathbf{V}^{k}{\mathbf{q}}(t_j),
\mathbf{V}^{k}\dot{{\mathbf{q}}}(t_j),
\mathbf{V}_{z}{\mathbf{q}}_{z}(t_j);
\boldsymbol{\gamma}
\right)
\right\|^2 .
\end{equation}
The total loss is then augmented by
$\beta_{z}\mathcal{L}_{GE}^{(z)}$. This joint reduction of structural responses and internal variables allows known constitutive evolution laws, such as Bouc--Wen-type hysteretic models, to be incorporated directly within
the reduced-order physics-informed formulation.

The data term penalizes the discrepancy between the model prediction and available sensor measurements (e.g., displacement, velocity, or acceleration), which arises from both measurement noise and reduced-order projection error. For acceleration measurements, the observation model can be written as follows:
\begin{equation}\label{eq:meas_acc}
\mathbf{y}_{\mathrm{meas}}(t)
=
\mathbf{S}_m \ddot{\mathbf{u}}(t) + \boldsymbol{\eta}(t)
=
\mathbf{S}_m \big( \mathbf{V}^{k}\ddot{\mathbf{q}}(t) + \ddot{\boldsymbol{\varepsilon}}^{k}(t) \big)
+
\boldsymbol{\eta}(t),
\end{equation}
where $\mathbf{S}_m$ is the selection matrix mapping model DOFs to sensor locations, $\boldsymbol{\eta}(t)$ denotes measurement noise, and $\boldsymbol{\varepsilon}^{k}(t)$ represents the error induced by the reduced basis. The reduced acceleration $\ddot{\mathbf{q}}(t)$ is obtained via automatic differentiation of the network output with respect to time.

Accordingly, the data loss is defined as follows:
\begin{equation}\label{eq:data_loss}
\mathcal{L}_S
=
\frac{1}{N_s}\sum_{j=1}^{N_s}
\left\|
\mathbf{S}_m \ddot{\boldsymbol{\varepsilon}}^{k}(t_j)
+
\boldsymbol{\eta}(t_j)
\right\|^2
=
\frac{1}{N_s}\sum_{j=1}^{N_s}
\left\|
\mathbf{y}_{\mathrm{meas}}(t_j)
-
\mathbf{S}_m \mathbf{V}^{k}
\ddot{{\mathbf{q}}}(t_j)
\right\|^2 .
\end{equation}
where $\{t_j\}_{j=1}^{N_s}$ are the measurement time instants. This loss penalizes the combined influence of measurement noise and errors induced by an imperfect reduced basis. The latter is not explicitly parameterized, but is reduced indirectly through physics-informed training and adaptive basis refinement.

\subsection{Basis Refinement for Physics Consistency}
\label{s3.3}

The reduced basis $\mathbf{V}$ defines the subspace in which the governing dynamics are enforced. As the inverse estimates evolve, a basis constructed from an initial or nominal model may become inconsistent with the identified system, increasing the unresolved component $\boldsymbol{\varepsilon}^{k}$ and the associated reduced consistency contribution ${\varepsilon}_{r}^{k}$. In residual-force identification, such basis mismatch may also cause the inferred residual force to partially compensate for reduced-order error rather than representing only the missing physical contribution.

To mitigate this effect, the RO-PINN couples the inverse optimization with an outer basis-refinement loop. For a fixed basis $\mathbf{V}^{k}$, the trainable variables are first optimized using the reduced-order loss described in Section~\ref{s3.2}. The resulting inverse estimates are then introduced into the full-order model in Eq.~\eqref{eq:basis_construction_k} to regenerate response snapshots. These snapshots are compressed using POD/SVD to construct the updated basis $\mathbf{V}^{k+1}$, after which the reduced operators and governing equations are re-projected onto the updated subspace.

In the present numerical implementation, the basis is recomputed directly from full-order response snapshots. For larger-scale applications, this step may be accelerated through local reduced-basis libraries and basis selection~\cite{amsallem2016pebl}, interpolation of parameter-dependent bases~\cite{vlachas2024parametric,vlachas2025reduced}, or hyper-reduction and matrix-compression strategies~\cite{agathos2024accelerating, balajewicz2016projection,kerfriden2013partitioned}.

The current basis is retained until either a gradient-based convergence condition is satisfied or the prescribed maximum number of inner optimization iterations is reached. The gradient-based criterion is
\begin{equation}
\label{eq:update_condition}
N_{\mathrm{iter}}^{k} \ge N_{\min},
\qquad
\left\|
\nabla_{\boldsymbol{\psi}} \mathcal{L}
\right\|_2
< \tau_g ,
\end{equation}
where $N_{\mathrm{iter}}^{k}$ is the number of inner optimization iterations performed under $\mathbf{V}^{k}$, $N_{\min}$ is the minimum number of iterations before convergence is assessed, $\tau_g$ is the prescribed global-gradient tolerance, and $\boldsymbol{\psi}$ collects the trainable variables active in the current inverse setting.

When Eq.~\eqref{eq:update_condition} is satisfied, the current inner optimization is terminated and the basis is updated. If the criterion is not reached within the prescribed maximum number of iterations $N_{\max}$, the basis is updated using the latest inverse estimates at $N_{\max}$. This allows earlier refinement after sufficient inner-loop convergence while limiting the
computational effort spent on any individual basis.

\subsection{Optimization Process}
\label{s3.4}

The RO-PINN minimizes the loss function in Eq.~\eqref{eq:total_loss} with respect to the trainable variables
\begin{equation}
\hat{\boldsymbol{\psi}}
=
\arg\min_{\boldsymbol{\psi}}
\mathcal{L}(\boldsymbol{\psi},\boldsymbol{\beta}),
\end{equation}
where $\boldsymbol{\psi}$ always includes the neural-network parameters $\boldsymbol{\theta}$ and additionally includes the structural parameters $\boldsymbol{\gamma}$ when parameter identification is performed. The residual force, when identified, is a neural-network output controlled by $\boldsymbol{\theta}$ rather than an independent trainable variable.

Training uses sensor measurements for the data loss and temporal collocation points for enforcing the reduced governing equations. For a fixed basis $\mathbf{V}^{k}$, the trainable variables are optimized using Adam~\cite{kingma2014adam} until either the gradient-based criterion in Eq.~\eqref{eq:update_condition} is satisfied or the maximum number of inner iterations $N_{\max}$ is reached. An optional L-BFGS-B stage ~\cite{byrd1995limited} may subsequently be applied for further convergence.

After each inner optimization stage, the current inverse estimates are passed to the basis-construction procedure described in Section~\ref{s3.3} to generate the updated basis $\mathbf{V}^{k+1}$. The reduced operators are then re-projected onto the updated subspace and the optimization continues. After final convergence, the full-field structural response is reconstructed from the optimized reduced coordinates as $\hat{\mathbf{u}}(t)=\mathbf{V}^{k}\hat{\mathbf{q}}(t)$, with velocity and acceleration obtained through automatic differentiation. When internal variables are included, their corresponding responses are reconstructed analogously from $\hat{\mathbf{z}}(t)=\mathbf{V}_{z}\hat{\mathbf{q}}_{z}(t)$. The complete procedure is summarized in Algorithm~\ref{alg:RO-PINN}.

\begin{algorithm}[h!]
\caption{Iterative training procedure for the RO-PINN framework with adaptive basis refinement}
\label{alg:RO-PINN}
\begin{algorithmic}[1]

\State \textbf{Initialize:} outer index $k=0$, network parameters
$\boldsymbol{\theta}$, structural parameters $\boldsymbol{\gamma}^{0}$ when
applicable, and initial basis $\mathbf{V}^{0}$.

\Repeat
    \State Sample temporal collocation points and sensor measurements.
    \State Train with fixed $\mathbf{V}^{k}$ using Adam.

    \If{
    $\left(
    N_{\mathrm{iter}}^{k} \ge N_{\min}
    \ \textbf{and}\
    \|\nabla_{\boldsymbol{\psi}}\mathcal{L}\|_2 < \tau_g
    \right)$
    \textbf{or}
    $N_{\mathrm{iter}}^{k} \ge N_{\max}$
    }
        \State Optionally apply L-BFGS-B for further convergence.
        \State Obtain the current inverse estimates.
        \State Generate full-order response snapshots using
        Eq.~\eqref{eq:basis_construction_k}.
        \State Construct $\mathbf{V}^{k+1}$ through POD/SVD and re-project
        the reduced operators.
        \State Set $k\leftarrow k+1$.
    \Else
        \State Keep $\mathbf{V}^{k}$ and continue optimization.
    \EndIf

\Until{final convergence}

\State \Return optimized variables $\hat{\boldsymbol{\psi}}$, final basis
$\mathbf{V}^{k}$, and reconstructed full-field response.

\end{algorithmic}
\end{algorithm}

%%%%%%%%%%%%%%%%%%%%%%%%%%%%%%%%%%%%%%%%%%%%%%%%%%%%%%%%%%%%%%%%%%%%%%
\section{Numerical Examples}
\label{s4}

\subsection{Four-Story Frame with Nonlinear Braces}
\label{s4.1}

\begin{figure}[h!]
    \centering
    \subfigure[Frame~A: nonlinear braces at all stories]{%
        \includegraphics[width=0.45\textwidth]{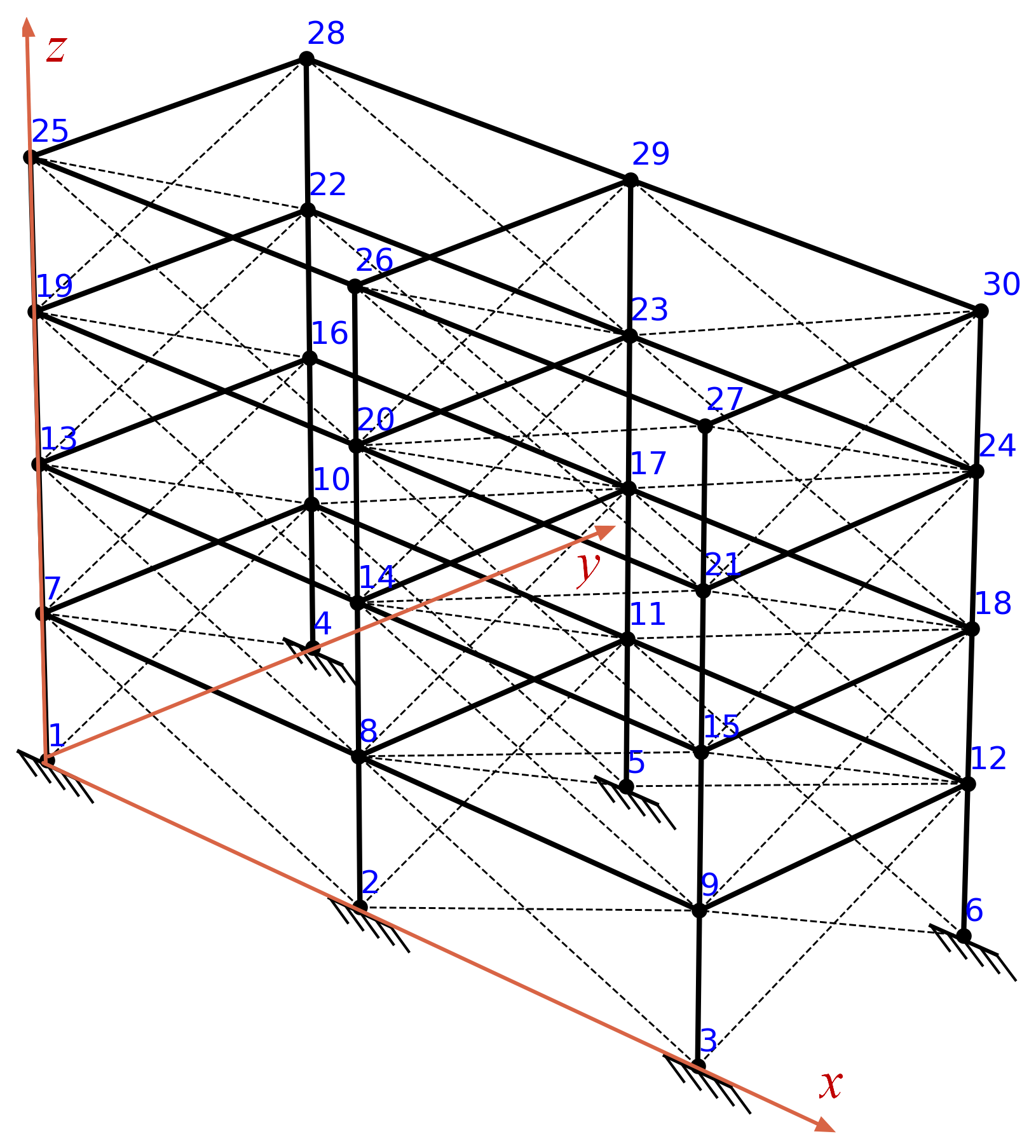}
    }\hspace{0.05\textwidth}
    \subfigure[Frame~B: nonlinear braces only at the first story]{%
        \includegraphics[width=0.45\textwidth]{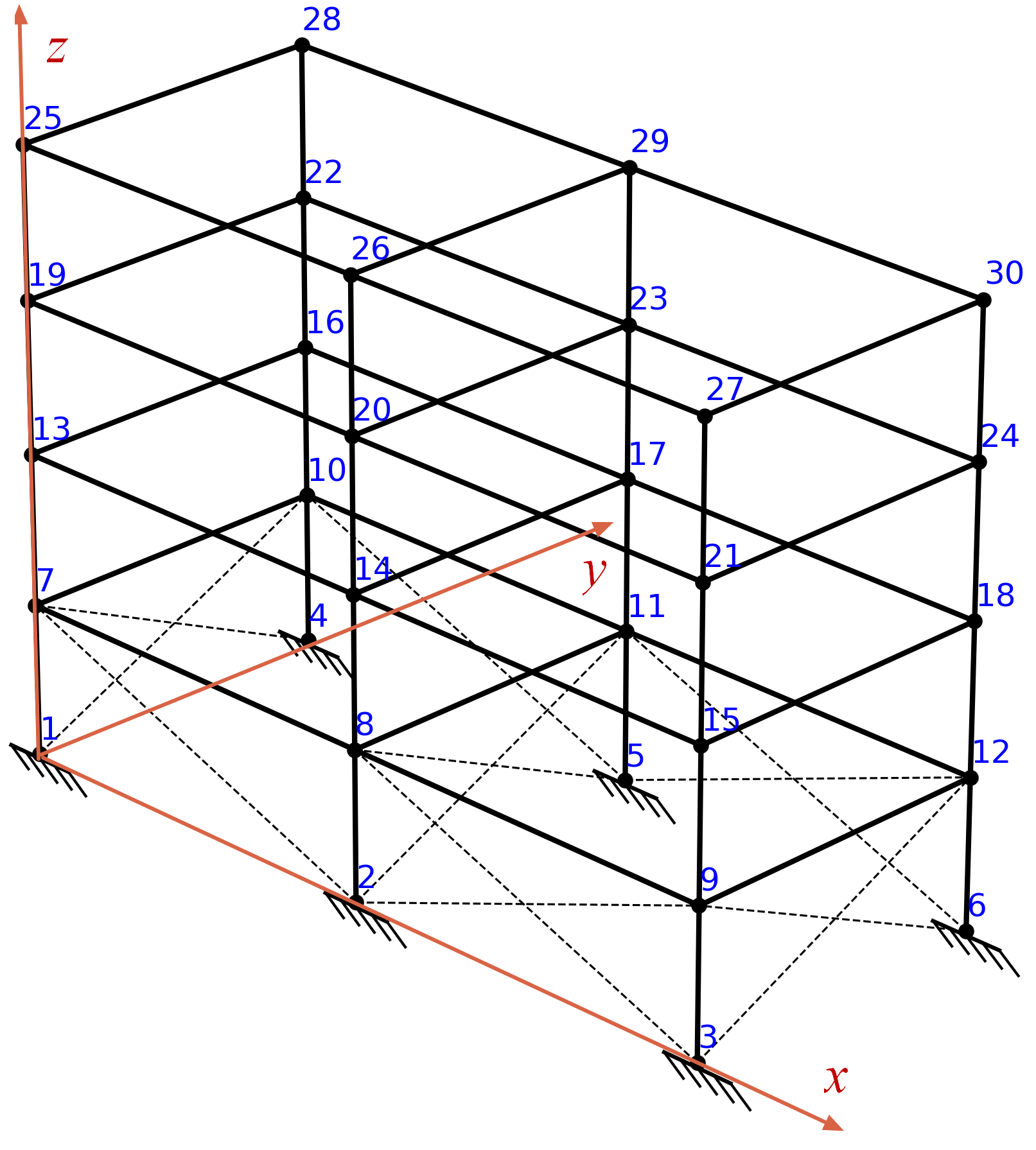}
    }
    \caption{Four-story two-bay steel frame benchmark with Bouc--Wen hysteretic
    braces: (a) Frame~A, with nonlinear braces at all stories; and
    (b) Frame~B, with nonlinear braces only at the first story, representing
    localized nonlinearity.}
    \label{eg01_frame}
\end{figure}

An open-source nonlinear three-dimensional shear-frame template
from~\cite{vlachas2021two} is adopted and extended to model a four-story, two-bay steel frame with nonlinear hysteretic braces, as illustrated in Figure~\ref{eg01_frame}. The structure consists of 30 nodes and 72 beam--column elements. Two configurations are considered: Frame~A, in which all stories are equipped with nonlinear braces, and Frame~B, in which only the first story is braced to represent localized nonlinearity. The braces follow a Bouc--Wen hysteretic model, which introduces nonlinear energy-dissipating behavior through internal variables.

The restoring force of the braces is expressed as
\begin{equation}
\mathbf{R}_{\mathrm{BW}}
=
\alpha k\,\delta\mathbf{u}
+
(1-\alpha)k\,\mathbf{z},
\end{equation}
where $\mathbf{R}_{\mathrm{BW}}$ denotes the Bouc--Wen brace restoring force, $\delta\mathbf{u}$ is the brace relative displacement, $\mathbf{z}$ is the hysteretic internal variable, $k$ is the initial stiffness, and $\alpha$ denotes the ratio of post-yield to initial stiffness, taken as $0.5$. The internal variable evolves according to
\begin{equation}
\dot{\mathbf{z}}
=
A\,\delta\dot{\mathbf{u}}
-
\beta
\left|\delta\dot{\mathbf{u}}\right|
|\mathbf{z}|^{n-1}\mathbf{z}
-
\gamma_{\mathrm{BW}}
\,\delta\dot{\mathbf{u}}
|\mathbf{z}|^{n},
\end{equation}
where $A$, $\beta$, $\gamma_{\mathrm{BW}}$, and $n$ are the Bouc--Wen parameters, set to $1$, $1$, $0.5$, and $2$, respectively. This formulation provides a nonlinear hysteretic benchmark for evaluating the reduced representation of structural and internal-variable dynamics and the adaptive basis-refinement strategy introduced in Section~\ref{s3}.

The cross-sectional and material properties of the columns, beams, and braces are summarized in Table~\ref{tab:member_props}. These properties define the reference configuration used in the subsequent numerical studies.

\begin{table}[htbp]
\centering
\caption{Geometric and mechanical properties of the four-story steel frame benchmark.}
\label{tab:member_props}
\begin{tabular}{p{5cm} p{5.5cm} p{5cm}}
\toprule
Geometric configuration & Material parameters & Cross-section \\
\midrule
Frame length (m): 6.4
& Young's modulus (GPa): 210
& Columns: HEA 200 \\
Frame width (m): 3.5
& Poisson's ratio: 0.3
& Beams: HEA 200 \\
Story height (m): 3.2
& Density (kg/m$^{3}$): 8000
& Braces: SHS 40$\times$40$\times$3 \\
\bottomrule
\end{tabular}
\end{table}

\subsection{Case Scenarios and Data Generation}
\label{s4.2}

Five representative cases are designed to evaluate the RO-PINN framework under three inverse settings: (i) residual-force identification under incomplete physics (Cases~C1--C2), in which the structure remains healthy but the nonlinear brace behavior is intentionally omitted from the assumed model; (ii) parameter identification under known physics (Cases~C3--C4), where structural damage is represented through stiffness-reduction factors; and (iii) joint identification of residual restoring forces and structural parameters acting on different stories (Case~C5). The configurations and corresponding unknown quantities are summarized in Table~\ref{tab:case_summary}.

\begin{table}[htbp]
\centering
\caption{Summary of simulation cases and corresponding unknown quantities.}
\label{tab:case_summary}
\begin{tabular}{p{2.5cm}p{8cm}p{4.5cm}}
\toprule
Case & Physics condition & Unknowns \\
\midrule
C1 (Frame~A)
& Healthy structure; missing nonlinear brace behavior at all stories
& $\mathbf{R}_r$ \\

C2 (Frame~B)
& Healthy structure; missing nonlinear brace behavior at the first story
& $\mathbf{R}_r$ \\

C3 (Frame~A)
& Damaged frame and braces at all stories
& $\boldsymbol{\gamma}
=\{k_1 ~\text{to} ~k_4,k_{b1} ~\text{to} ~k_{b4}\}$ \\

C4 (Frame~B)
& Healthy frame; localized damage in first-story braces
& $\boldsymbol{\gamma}
=\{k_{b1}^{(1)},\ldots,k_{b1}^{(12)}\}$ \\

C5 (Frame~B)
& Missing nonlinear brace behavior at the first story and interstory
stiffness damage at the upper three stories
& $\{\mathbf{R},\,
\boldsymbol{\gamma}=\{k_2,k_3,k_4\}\}$ \\
\bottomrule
\end{tabular}
\end{table}

In Cases~C1--C2, all frame and brace components remain healthy, while the assumed model intentionally excludes the nonlinear brace contribution. The resulting model-form error is represented through the learned reduced residual force $\mathbf{R}_r$. For basis refinement, the corresponding full-order residual-force representation is recovered and introduced into the full-order simulation used for snapshot generation.

In Cases~C3--C4, stiffness-reduction factors are introduced to represent structural degradation. A factor equal to one corresponds to the healthy state, whereas values below one indicate stiffness loss. For Case~C3 (Frame~A), the ground-truth interstory stiffness factors are $\{k_1,k_2,k_3,k_4\}=\{1.0,\,0.5,\,0.3,\,1.0\},$ while the brace stiffness factors are $\{k_{b1},k_{b2},k_{b3},k_{b4}\}=\{0.4,\,1.0,\,1.0,\,0.2\}$. For Case~C4 (Frame~B), the frame remains healthy and damage is confined to the twelve first-story braces as $\{k_{b1}^{(1)},\ldots,k_{b1}^{(12)}\}=\{0.4\times3,\,1.0\times7,\,0.2\times2\}$.

Case~C5 combines residual-force and parameter identification within a single inverse problem while spatially separating the two classes of unknowns. The nonlinear restoring forces of the twelve first-story braces are treated as unknown physical-space residual-force components, whereas the stiffness factors of the upper three stories, $\boldsymbol{\gamma}=\{k_2,k_3,k_4\}$, are identified simultaneously as trainable variables. Their ground-truth values are $\{k_2,k_3,k_4\}=\{0.9,\,0.6,\,1.0\}$, while the first-story frame remains healthy.

The initial reduced bases are constructed from the healthy configuration. For Cases~C1--C2, the initial model excludes the nonlinear brace contribution, whereas Cases~C3--C4 retain the known nonlinear brace model with nominal parameters. Case~C5 starts from the corresponding healthy linear-frame configuration and uses the additional static enrichment described below.

Each scenario is simulated using the Newmark--$\beta$ method under the El~Centro earthquake excitation. The ground motion is normalized to a peak ground acceleration of $0.2\,g$ and applied horizontally at the base in both the $x$- and $y$-directions for a total duration of $10\,\mathrm{s}$, discretized into 1000 time steps ($\Delta t=0.01\,\mathrm{s}$). Rayleigh damping is selected to reproduce a $2\%$ critical damping ratio at the first natural frequency ($f_1=2.3$~Hz). Synthetic displacement and acceleration measurements are collected at selected floor nodes and used in the data loss $\mathcal{L}_S$ of Eq.~\eqref{eq:data_loss}. Two sensors are placed at corner nodes on each floor. At the first floor, displacement measurements are available at nodes 7 and 12, while acceleration measurements are provided at the corresponding corner nodes of the upper three floors.

To emulate measurement and modeling uncertainty, random $2\%$ variations are introduced in selected material properties and $5\%$ Gaussian noise is added to the measured responses. The initial dynamic basis $\mathbf{V}^{0}$ is obtained through POD/SVD of simulated displacement snapshots under the same excitation. The first three modes are retained, capturing more than $99.99\%$ of the squared singular-value content of the snapshot matrix.

For Case~C5, the three-mode dynamic POD basis is augmented with four static-correction vectors to improve the representation of localized first-story force effects. Static flexibility responses associated with the twelve first-story brace-force patterns are first computed as
\begin{equation}
\mathbf{X}_{s}
=
\mathbf{K}_{\mathrm{health}}^{-1}\mathbf{S}_{f}^{\top},
\end{equation}
where $\mathbf{S}_{f}$ maps the twelve brace-force components to the full-order degrees of freedom. These flexibility responses are compressed by SVD, $\mathbf{X}_{s}=\mathbf{U}_{s}\boldsymbol{\Sigma}_{s}\mathbf{W}_{s}^{\top}$, and the first four left singular vectors are retained as the static-correction subspace. Following the classical static-correction concept~\cite{dickens1997critique}, this enrichment introduces localized deformation patterns that are poorly represented by the global POD modes.

The underlying static-correction subspace is computed once from the nominal healthy model. After each dynamic-basis update, the retained static vectors are deflated against the updated POD basis and re-orthonormalized before being appended. The Case~C5 reduced basis therefore consists of three dynamically updated POD modes and four static-correction vectors, giving a total reduced dimension of seven.

\subsection{Training Implementation and Evaluation}
\label{s4.3}

All cases use the same fully connected feed-forward network architecture, consisting of four hidden layers with 64 neurons per layer and $\tanh$ activation functions. The network weights are initialized using Xavier initialization~\cite{glorot2010understanding}. Time $t$ is used as the network input, and the output always includes the reduced coordinates $\mathbf{q}(t)$. Additional outputs are included according to the inverse setting, such as reduced internal variables and/or residual-force representations as described in Section~\ref{s3.2}. For Case~C5, the residual force is represented directly by the unknown first-story brace-force components, while the upper-story stiffness factors are treated as additional trainable variables.

Training follows the iterative procedure described in Section~\ref{s3.4}. For a fixed basis $\mathbf{V}^{k}$, the trainable variables are optimized using Adam~\cite{kingma2014adam} with a learning rate of $10^{-3}$. The inner optimization is terminated when the gradient-based criterion in Eq.~\eqref{eq:update_condition} is satisfied or when the prescribed maximum number of iterations $N_{\max}$ is reached. An optional L-BFGS-B stage~\cite{byrd1995limited} may subsequently be applied for further loss minimization.

After each inner optimization stage, the current inverse estimates are used to regenerate full-order response snapshots and construct the updated dynamic basis through POD. For Cases~C3--C4, the full-order simulation incorporates the current stiffness-reduction factors, whereas for Cases~C1--C2 it incorporates the current residual-force estimate. Case~C5 follows the same basis-refinement procedure using the simultaneously identified upper-story stiffness factors and first-story residual brace forces, with the dynamic basis additionally enriched by the four static-correction vectors described in Section~\ref{s4.2}. The reduced operators are re-projected after each basis update.

All experiments are implemented in Python/TensorFlow~\cite{abadi2016tensorflow} and executed on a workstation equipped with an
Intel Core~i7 CPU, 64~GB of RAM, and an NVIDIA RTX4090 GPU with 24~GB of memory. Performance is evaluated using the normalized $\ell_2$ error,
\begin{equation}
\text{Error}~(\%)
=
\frac{
\left\|
\mathbf{y}_{\mathrm{true}}
-
\mathbf{y}_{\mathrm{pred}}
\right\|_2
}{
\left\|
\mathbf{y}_{\mathrm{true}}
\right\|_2
}
\times 100 ,
\end{equation}
where $\mathbf{y}$ denotes the quantity being evaluated, including identified parameters, residual forces, or structural responses. Computational efficiency is assessed using the reported total training time and, where relevant, the average computational cost per optimization iteration.

\subsection{Results and Analysis}
\label{s4.4}

The performance of the proposed RO-PINN framework under the five cases defined in Section~\ref{s4.2} is evaluated in terms of identification accuracy, response reconstruction, and computational efficiency. The results are organized according to the three inverse settings considered in this work: (i) residual-force identification under incomplete physics (Cases~C1--C2), (ii) parameter identification under known physics (Cases~C3--C4), and (iii) joint residual-force and parameter identification (Case~C5). The analysis focuses on the recovery of the unknown quantities, reconstruction of structural responses at instrumented and uninstrumented degrees of freedom, and the influence of adaptive basis refinement.

\subsubsection{Model-Form Error Correction through Residual-Force Identification}
\label{s4.4.1}

Cases~C1 (Frame~A) and C2 (Frame~B) evaluate the ability of the RO-PINN framework to correct model-form error through residual-force identification under incomplete physics. In both cases, the structures remain healthy, while the nonlinear brace contribution is intentionally omitted from the assumed governing equations. Parameter-only model updating cannot represent this missing contribution because no corresponding parameter is introduced within the assumed model. The RO-PINN instead learns a reduced residual-force representation while adaptively refining the reduced basis to reduce basis-mismatch effects and improve consistency with the corrected dynamics.

\begin{figure}[h!]
    \centering
    \subfigure[Mode 1]{
        \begin{minipage}[b]{0.9\textwidth}
            \includegraphics[width=1\textwidth]{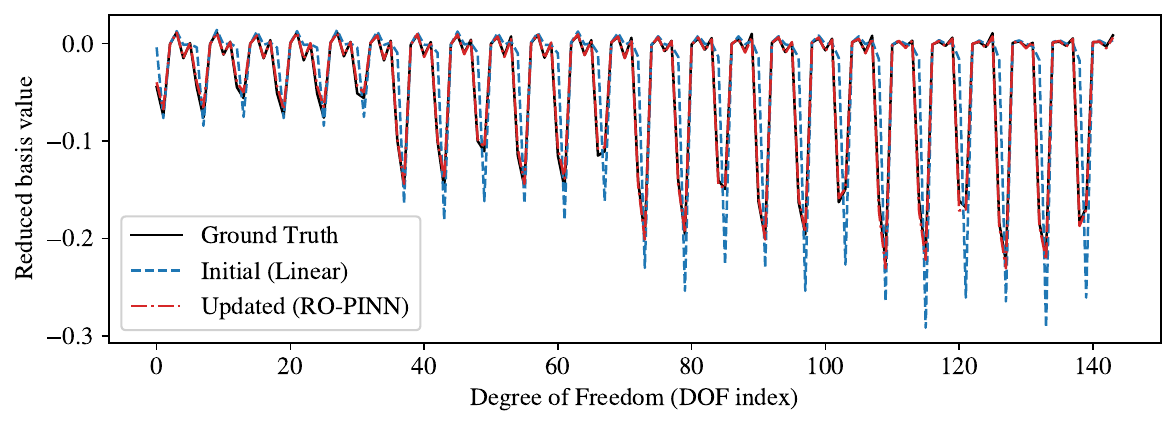}
        \end{minipage}}
    \subfigure[Mode 2]{
        \begin{minipage}[b]{0.9\textwidth}
            \includegraphics[width=1\textwidth]{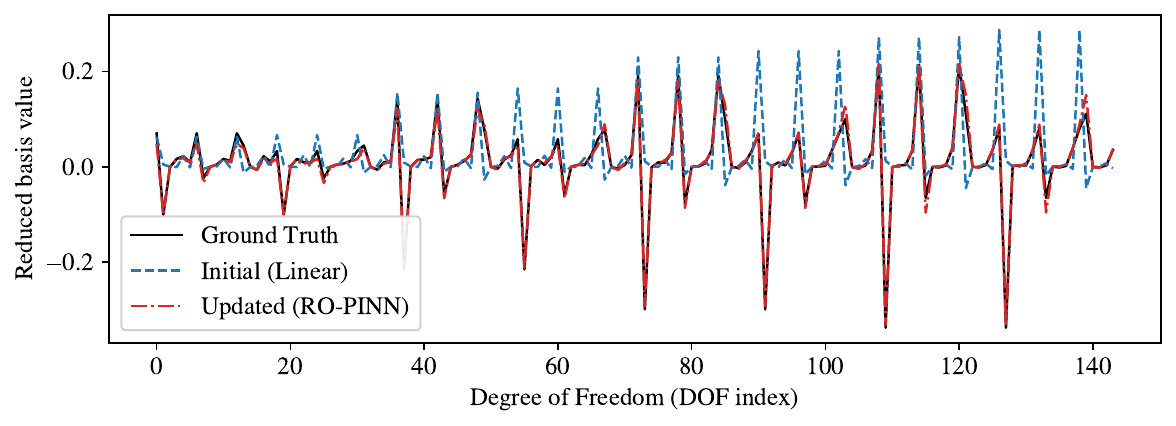}
        \end{minipage}}
    \subfigure[Mode 3]{
        \begin{minipage}[b]{0.9\textwidth}
            \includegraphics[width=1\textwidth]{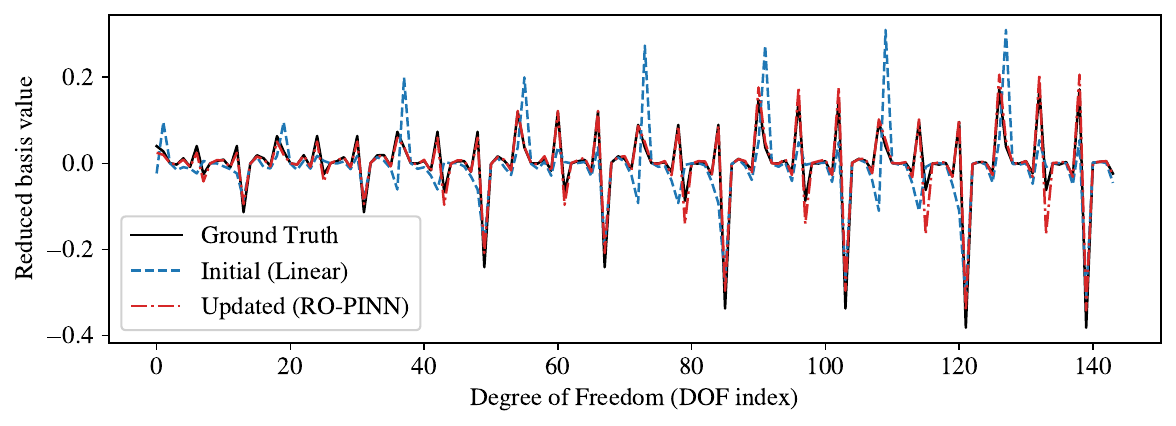}
        \end{minipage}}
    \caption{Comparison of reduced bases for Case~C1: initial basis constructed
    from the nominal linear model, adaptively refined basis obtained by RO-PINN,
    and reference basis extracted from the complete nonlinear model, shown for
    the first three modes.}
    \label{fig:mfe_basis}
\end{figure}

Figure~\ref{fig:mfe_basis} compares the initial basis constructed from the nominal linear model, the basis obtained after adaptive refinement, and the reference basis extracted from the complete nonlinear model for Case~C1. The initial basis reflects the nominal linear dynamics and therefore differs from the basis associated with the nonlinear system. After refinement, the updated basis more closely approaches the reference nonlinear basis, indicating that the reduced subspace progressively adapts to the corrected system dynamics. This reduces the contribution of basis mismatch to the reduced equilibrium residual and limits its influence on the inferred residual force.

\begin{figure}[h!]
    \centering
    \subfigure[$t$ vs.\ $\mathbf{R}$ using initial basis]{
        \begin{minipage}[b]{0.48\textwidth}
            \includegraphics[width=1\textwidth]{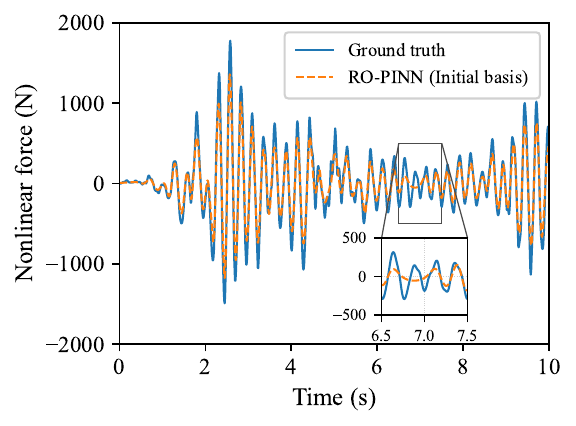}
        \end{minipage}}
    \subfigure[$t$ vs.\ $\mathbf{R}$ using adaptively refined basis]{
        \begin{minipage}[b]{0.48\textwidth}
            \includegraphics[width=1\textwidth]{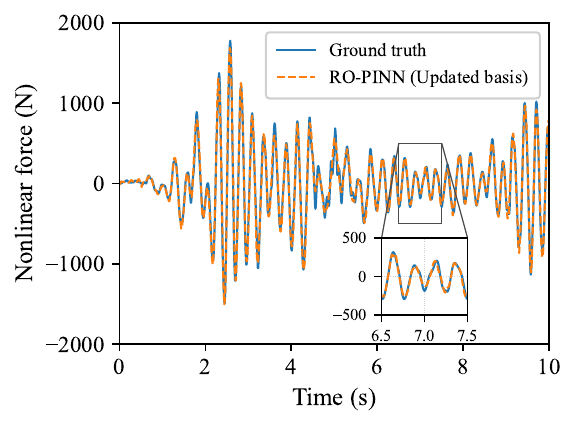}
        \end{minipage}}
    \caption{Learned residual force for Case~C1: comparison between the
    reconstructed RO-PINN residual force $\mathbf{R}$ and the reference
    nonlinear brace restoring force using (a) the initial basis and
    (b) the adaptively refined basis.}
    \label{fig:mfe_residual}
\end{figure}

Figure~\ref{fig:mfe_residual} compares the reconstructed full-order residual force obtained from the learned reduced residual $\mathbf{R}_r$ with the reference nonlinear brace restoring force for Case~C1. With the initial fixed basis, noticeable discrepancies remain in the identified force history. After adaptive basis refinement, the reconstructed residual force shows substantially closer agreement with the reference response, indicating that refinement of the reduced subspace improves the separation between the missing physical contribution and reduced-basis error.

Table~\ref{tab:mfe_errors} summarizes the normalized errors in displacement $u$, velocity $\dot{u}$, acceleration $\ddot{u}$, and nonlinear restoring force $R$ for Cases~C1 and C2. Adaptive basis refinement improves both state and force reconstruction, with the largest gains observed in the residual-force estimates. For Case~C1, the force error decreases from $29.4\%$ to $9.2\%$, while for Case~C2 it decreases from $38.3\%$ to $14.7\%$. The larger force error in Case~C2 is consistent with the greater difficulty of identifying a spatially localized nonlinear contribution from limited measurements.

\begin{table}[h!]
\centering
\caption{Model-form error correction cases C1--C2: normalized errors (\%) for
displacement $u$, velocity $\dot{u}$, acceleration $\ddot{u}$, and nonlinear
restoring force $R$, with and without adaptive basis refinement.}
\label{tab:mfe_errors}
\begin{tabular}{p{3cm} p{3cm} p{1.5cm} p{1.5cm} p{1.5cm} p{1.5cm}}
\toprule
Case & Basis updating & $u$ & $\dot{u}$ & $\ddot{u}$ & $R$ \\
\midrule
C1 (Frame~A) & Without & 4.8 & 5.1 & 5.0 & 29.4 \\
              & With    & 3.1 & 3.3 & 3.8 & 9.2 \\
C2 (Frame~B) & Without & 4.4 & 5.0 & 4.8 & 38.3 \\
              & With    & 3.3 & 3.5 & 4.2 & 14.7 \\
\bottomrule
\end{tabular}
\end{table}

\begin{figure}[h!]
    \centering
    \subfigure[Sensor node~30, $\ddot{u}_x$]{
        \begin{minipage}[b]{0.45\textwidth}
            \includegraphics[width=1\textwidth]{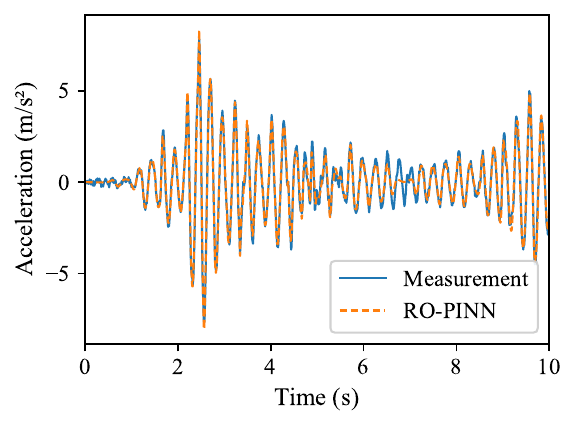}
        \end{minipage}}
    \subfigure[Sensor node~30, $\ddot{u}_y$]{
        \begin{minipage}[b]{0.45\textwidth}
            \includegraphics[width=1\textwidth]{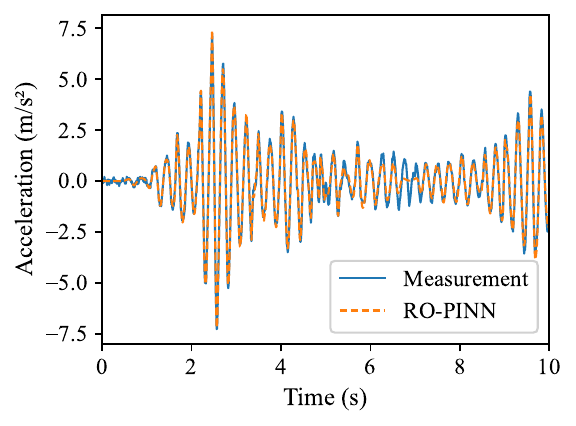}
        \end{minipage}}\\
    \subfigure[Non-sensor node~27, $u_x$]{
        \begin{minipage}[b]{0.45\textwidth}
            \includegraphics[width=1\textwidth]{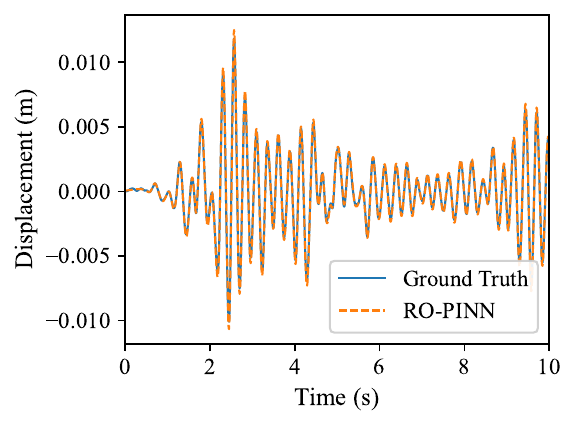}
        \end{minipage}}
    \subfigure[Non-sensor node~27, $u_y$]{
        \begin{minipage}[b]{0.45\textwidth}
            \includegraphics[width=1\textwidth]{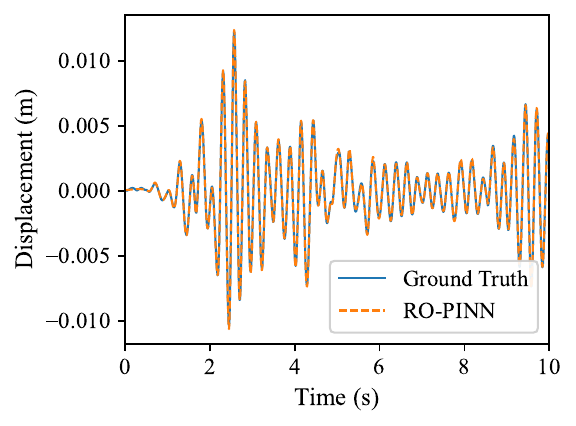}
        \end{minipage}}
    \subfigure[Non-sensor node~27, $\theta_x$]{
        \begin{minipage}[b]{0.45\textwidth}
            \includegraphics[width=1\textwidth]{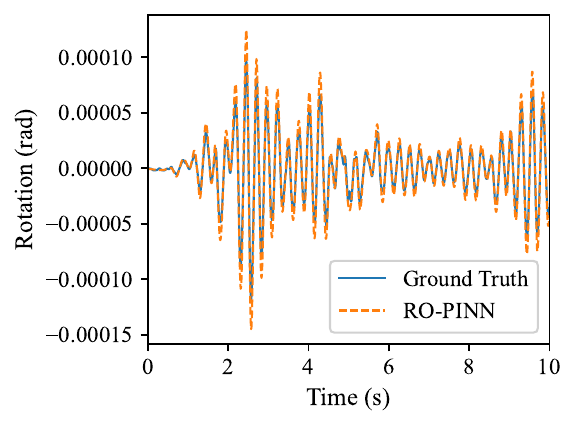}
        \end{minipage}}
    \subfigure[Non-sensor node~27, $\theta_y$]{
        \begin{minipage}[b]{0.45\textwidth}
            \includegraphics[width=1\textwidth]{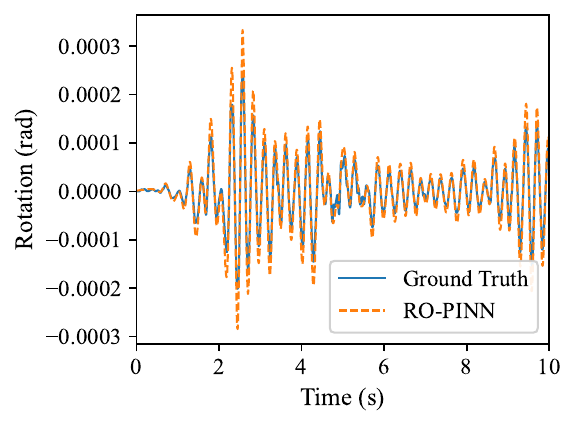}
        \end{minipage}}
    \caption{Response reconstruction for Case~C1: predicted and reference
    responses at instrumented node~30, including accelerations in the
    (a) $x$ and (b) $y$ directions, and at uninstrumented node~27, including
    translational displacements (c) $u_x$ and (d) $u_y$, and rotational
    responses (e) $\theta_x$ and (f) $\theta_y$.}
    \label{fig:mfe_resp}
\end{figure}

Finally, Figure~\ref{fig:mfe_resp} compares representative response reconstructions for Case~C1 at instrumented node~30 and uninstrumented node~27. The predicted responses closely follow the reference solution at both measured and unmeasured locations, indicating that the learned residual contribution, together with the reduced equilibrium constraints, enables reconstruction of the global structural response despite the omitted nonlinear brace physics.

\subsubsection{Parameter Identification}
\label{s4.4.2}

\begin{figure}[h!]
	\centering
	\subfigure[]{
		\begin{minipage}[b]{0.95\textwidth}
			\includegraphics[width=1\textwidth]{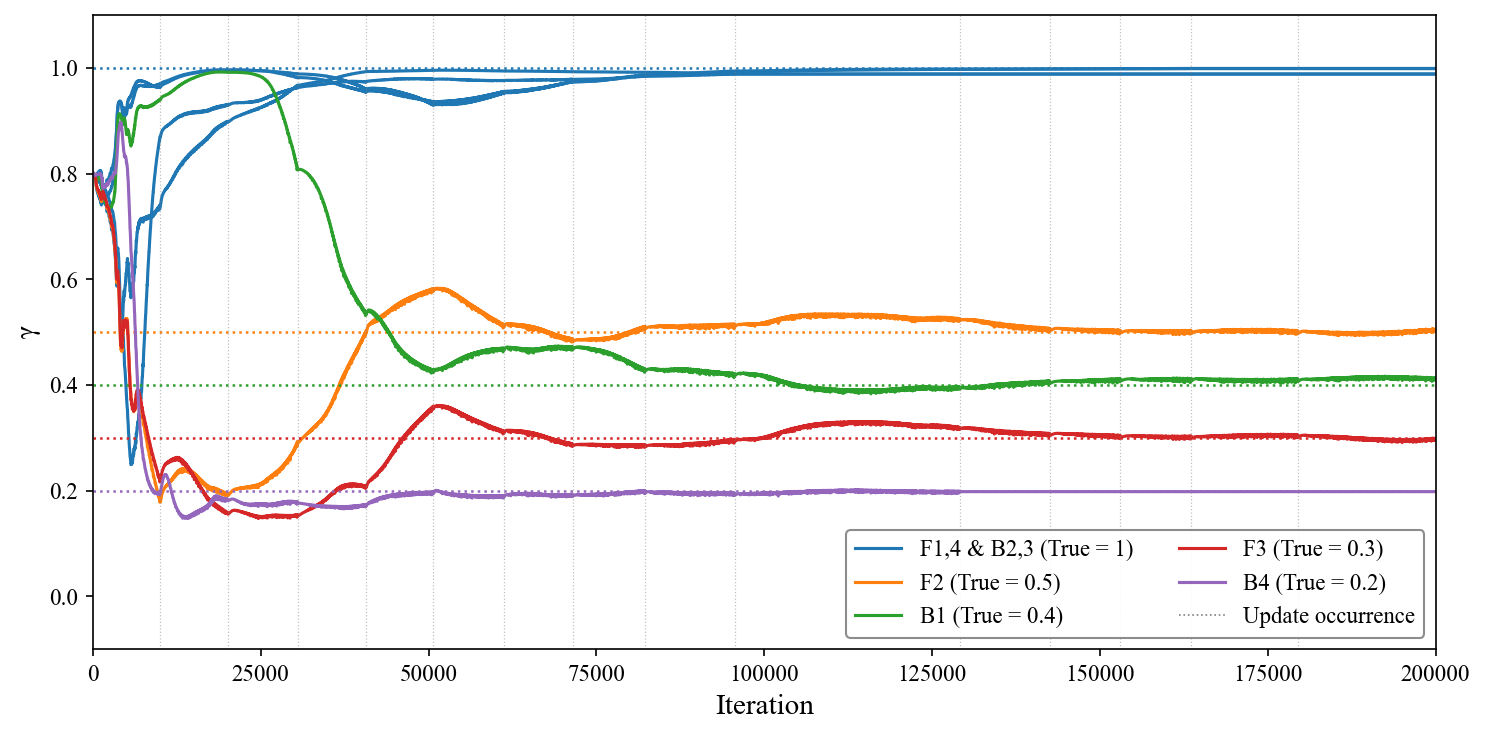}
		\end{minipage}
	}
	\subfigure[]{
		\begin{minipage}[b]{0.95\textwidth}
			\includegraphics[width=1\textwidth]{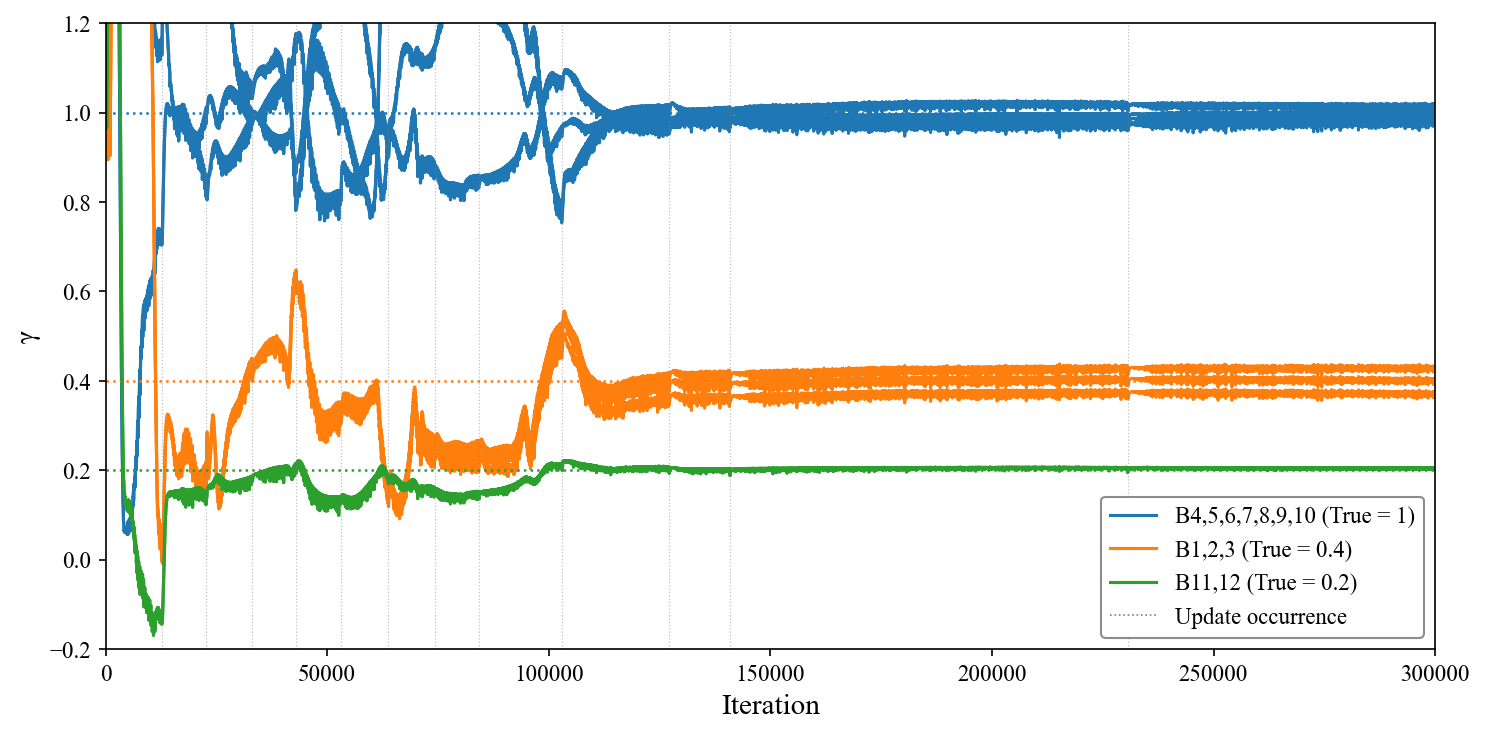}
		\end{minipage}
	}
	\caption{Convergence histories of the identified stiffness parameters for
	Cases~C3--C4. Solid lines denote RO-PINN estimates, dashed lines denote the
	ground-truth values, and vertical markers indicate basis-update events
	triggered according to Eq.~\eqref{eq:update_condition}.}
	\label{fig:params_evolution}
\end{figure}

Cases~C3--C4 evaluate parameter identification under known physics, where the functional form of the governing equations is prescribed but selected stiffness parameters are unknown. These parameters are treated as trainable variables and optimized together with the neural-network parameters. Figure~\ref{fig:params_evolution} shows their convergence histories. In both cases, the identified parameters progressively approach the ground-truth values as the optimization proceeds. The vertical markers indicate basis-update events triggered according to Eq.~\eqref{eq:update_condition}, through which the reduced operators are recomputed using the updated parameter estimates.

\begin{figure}[h!]
	\centering
	\subfigure[$t$ vs.\ $z$ using initial basis]{
		\begin{minipage}[b]{0.5\textwidth}
			\includegraphics[width=1\textwidth]{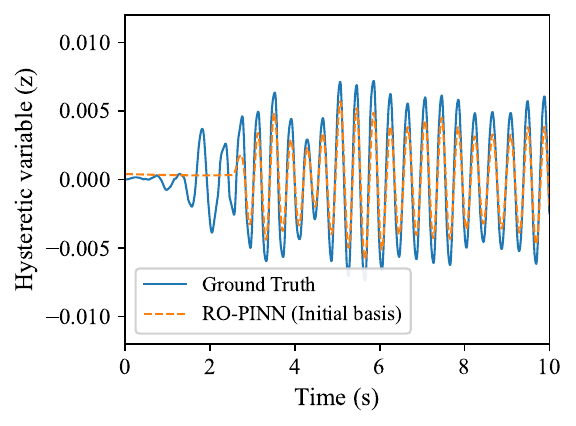}
		\end{minipage}
	}
	\subfigure[$\Delta u$ vs.\ $z$ using initial basis]{
		\begin{minipage}[b]{0.44\textwidth}
			\includegraphics[width=1\textwidth]{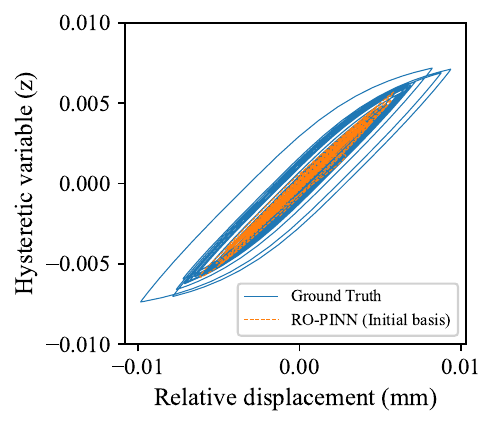}
		\end{minipage}
	}
	\subfigure[$t$ vs.\ $z$ using adaptively refined basis]{
		\begin{minipage}[b]{0.5\textwidth}
			\includegraphics[width=1\textwidth]{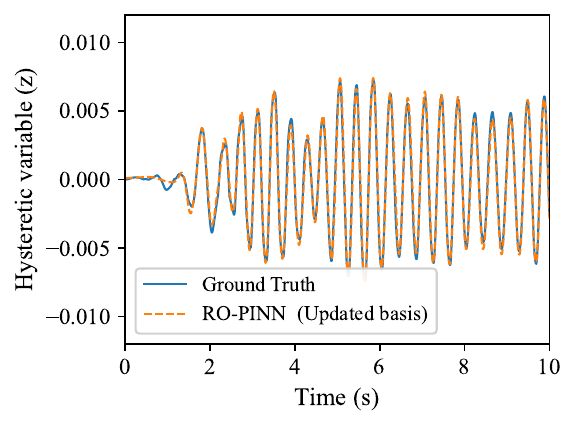}
		\end{minipage}
	}
	\subfigure[$\Delta u$ vs.\ $z$ using adaptively refined basis]{
		\begin{minipage}[b]{0.44\textwidth}
			\includegraphics[width=1\textwidth]{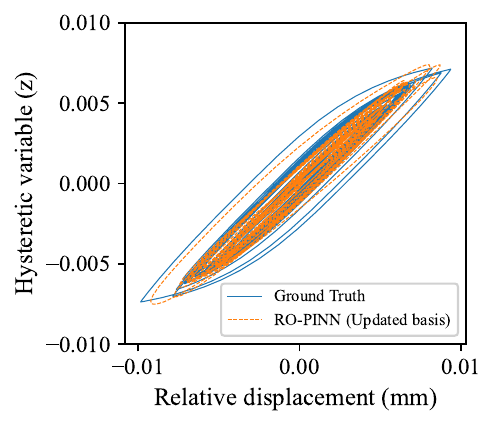}
		\end{minipage}
	}
	\caption{Reconstruction of the Bouc--Wen response for Case~C4 at a
	representative brace element (Nodes~3--12): comparison of RO-PINN predictions
	with the reference solution for (a,c) the internal-variable evolution and
	(b,d) the hysteretic relationship between brace deformation and internal
	variable $z$, using the initial and adaptively refined bases, respectively.}
	\label{fig:case1_brace}
\end{figure}

Figure~\ref{fig:case1_brace} compares reconstructed Bouc--Wen responses for a representative brace element in Case~C4. With the initial fixed basis, noticeable discrepancies remain in both the internal-variable evolution and the force--deformation relationship. After adaptive basis refinement, the predicted internal-variable history and hysteretic response show substantially closer agreement with the reference solution, including the principal loading and unloading branches. This indicates that updating the reduced subspace improves consistency between the evolving parameter estimates and the nonlinear reduced dynamics.

Table~\ref{tab:results} summarizes the parameter-estimation errors, state reconstruction errors, and computational cost for Cases~C3--C4. For a consistent comparison with BMU, the reported state errors are obtained by re-simulating the full-order model using the identified parameters and the Newmark--$\beta$ integrator. This separates parameter-estimation accuracy from reduced-order projection error. Direct state reconstruction from the trained RO-PINN, $\hat{\mathbf{u}}=\mathbf{V}\hat{\mathbf{q}}$, is reported separately in the table footnote.

Without basis refinement, the parameter estimates do not converge satisfactorily, with mean errors of $46.2\%$ for Case~C3 and exceeding $100\%$ for Case~C4. With adaptive basis refinement, the mean errors decrease to $1.77\%$ and $3.43\%$, respectively. The corresponding state and internal-variable errors also decrease substantially. These results show that updating the reduced basis as the parameter estimates evolve is important for maintaining an accurate reduced representation during the inverse solution.

\begin{table}[h!]	
	\caption{Parameter identification accuracy and computational cost for
	Cases~C3--C4, comparing RO-PINN results with and without adaptive basis
	refinement against Bayesian model updating (BMU). State errors for all
	methods are evaluated through full-order Newmark--$\beta$ re-simulation
	using the identified parameters.\textsuperscript{a}}
    \label{tab:results}
	\begin{minipage}{\textwidth}
		\centering
	    \footnotesize
		\renewcommand{\arraystretch}{1}
		\begin{tabular} {llcccccccc}
		\toprule
            \multirow{2}{0.5cm}{Case} & \multirow{2}{1.5cm}{Method}
            & \multicolumn{2}{c}{Parameter errors (\%)}
            & \multicolumn{3}{c}{State errors (\%)}
            & \multicolumn{2}{c}{Computational time}\\ 
		\cmidrule{3-9}
		   & & Mean & Max & $u$ & $\dot{u}$ & $z$
		   & Per iter. (s) & Total (h)\\ 
			\cmidrule{1-9}
			\multirow{6}{*}{C3}
			& RO-PINN & 46.2 & >100 & 14.2 & 15.0 & 35.7 & 0.01 & 1.5 \\ 
            & (no refinement) & & & & & & & \small{(no converg.)}\\
			\cmidrule{2-9}
			& RO-PINN & 1.77 & 5.06 & 1.30 & 1.46 & 2.34 & 0.01 & 1.0\\ 
            & (basis refinement) & & & & & & & \\
			\cmidrule{2-9}
			& BMU & 3.12 & 5.60 & 1.06 & 1.17 & 3.25 & 5.22 & 5.8 \\ 
            & (1000 samples) & & & & & & (per sample)
            & \small{(4 paral. cores)} \\ 
			\cmidrule{1-9}
			\multirow{6}{*}{C4}
			& RO-PINN & >100 & >100 & 27.9 & 26.7 & 35.8 & 0.01 & 1.5\\ 
            & (no refinement) & & & & & & & \small{(no converg.)}\\
			\cmidrule{2-9}
			& RO-PINN & 3.43 & 7.81 & 1.65 & 1.70 & 1.78 & 0.01 & 0.9\\ 
            & (basis refinement) & & & & & & & \\
            \cmidrule{2-9}
            & BMU & 4.50 & 6.95 & 1.85 & 2.37 & 5.15 & 5.05 & 5.4 \\ 
            & (1000 samples) & & & & & & (per sample)
            & \small{(4 paral. cores)} \\ 
			\bottomrule
		\end{tabular}
        \\[2pt]
        \raggedright\footnotesize
        \textsuperscript{a}~Direct reduced-order state reconstruction from the
        network output $\hat{\mathbf{u}}=\mathbf{V}\hat{\mathbf{q}}$ yields
        errors of approximately $5\%$ and $2\%$ for Cases~C3 and C4,
        respectively. These values additionally include reduced-order projection
        error, but require no additional forward simulation.
	\end{minipage}
\end{table}

\begin{figure}[h!]
	\centering
	\subfigure[Sensor node~30, $\ddot{u}_x$]{
		\begin{minipage}[b]{0.45\textwidth}
			\includegraphics[width=1\textwidth]{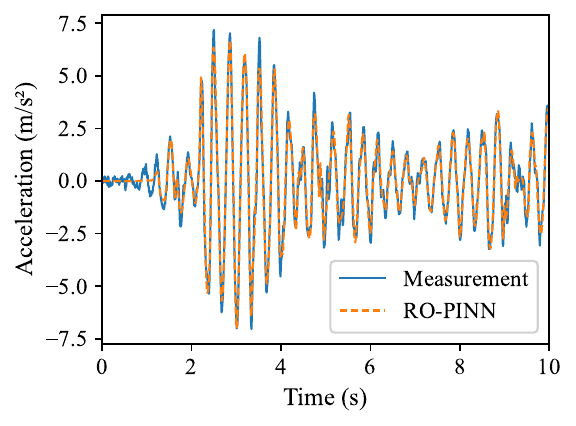}
		\end{minipage}
	}
	\subfigure[Sensor node~30, $\ddot{u}_y$]{
		\begin{minipage}[b]{0.45\textwidth}
			\includegraphics[width=1\textwidth]{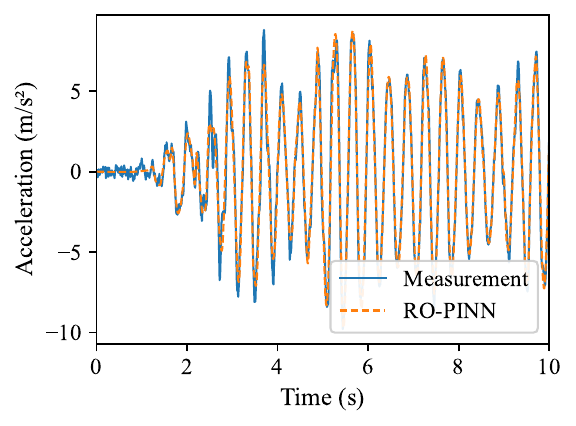}
		\end{minipage}
	}
	\subfigure[Non-sensor node~27, $u_x$]{
		\begin{minipage}[b]{0.45\textwidth}
			\includegraphics[width=1\textwidth]{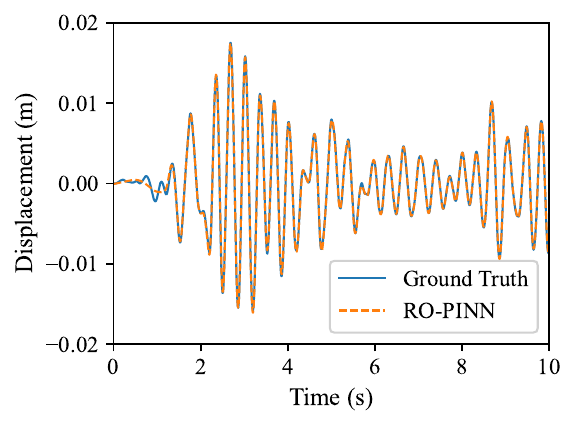}
		\end{minipage}
	}
	\subfigure[Non-sensor node~27, $u_y$]{
		\begin{minipage}[b]{0.45\textwidth}
			\includegraphics[width=1\textwidth]{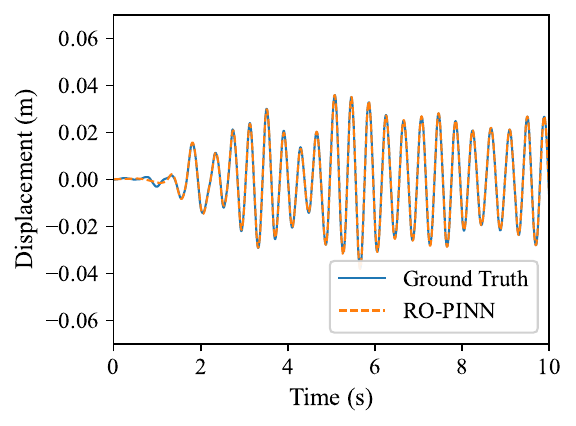}
		\end{minipage}
	}
	\subfigure[Non-sensor node~27, $\theta_x$]{
		\begin{minipage}[b]{0.45\textwidth}
			\includegraphics[width=1\textwidth]{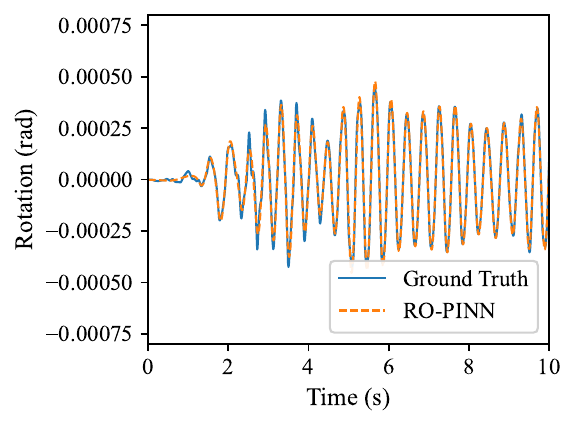}
		\end{minipage}
	}
	\subfigure[Non-sensor node~27, $\theta_y$]{
		\begin{minipage}[b]{0.45\textwidth}
			\includegraphics[width=1\textwidth]{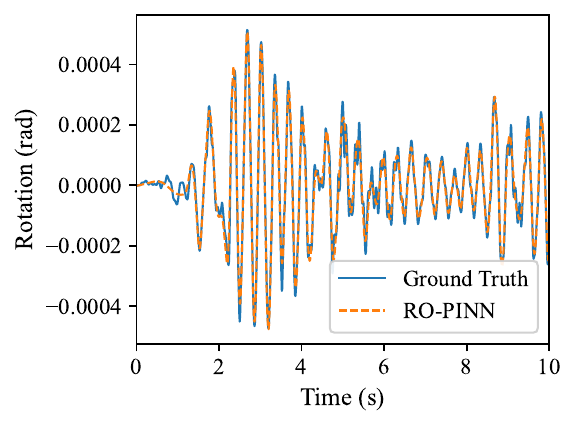}
		\end{minipage}
	}
    \caption{Response reconstruction for Case~C4: predicted and reference
    responses at instrumented node~30, including accelerations in the
    (a) $x$ and (b) $y$ directions, and at uninstrumented node~27, including
    translational displacements (c) $u_x$ and (d) $u_y$, and rotational
    responses (e) $\theta_x$ and (f) $\theta_y$.}
    \label{fig:case4_response}
\end{figure}

Figure~\ref{fig:case4_response} presents representative response reconstructions for Case~C4. The predicted acceleration responses at instrumented node~30 and the translational and rotational responses at uninstrumented node~27 closely follow the reference solution. Together with the parameter-identification results, this indicates that the refined reduced model can reconstruct the global structural response from sparse measurements after the underlying stiffness parameters have been identified.

\subsubsection{Joint Identification of Residual Force and Structural Parameters (Case~C5)}
\label{s4.4.3}

Case~C5 evaluates the RO-PINN framework in the joint setting, where the first-story residual restoring forces and the upper-story stiffness-reduction factors $\boldsymbol{\gamma}=\{k_2,k_3,k_4\}$ are identified simultaneously from the same sparse and noisy measurements. Unlike Cases~C1--C4, which consider residual-force identification and parameter identification separately, this case combines both sources of uncertainty within a single inverse problem. The same RO-PINN framework is used, with the additional static-correction enrichment described in Section~\ref{s4.2} to improve the representation of localized first-story force effects.

\begin{figure}[h!]
	\centering
	\includegraphics[width=0.95\textwidth]{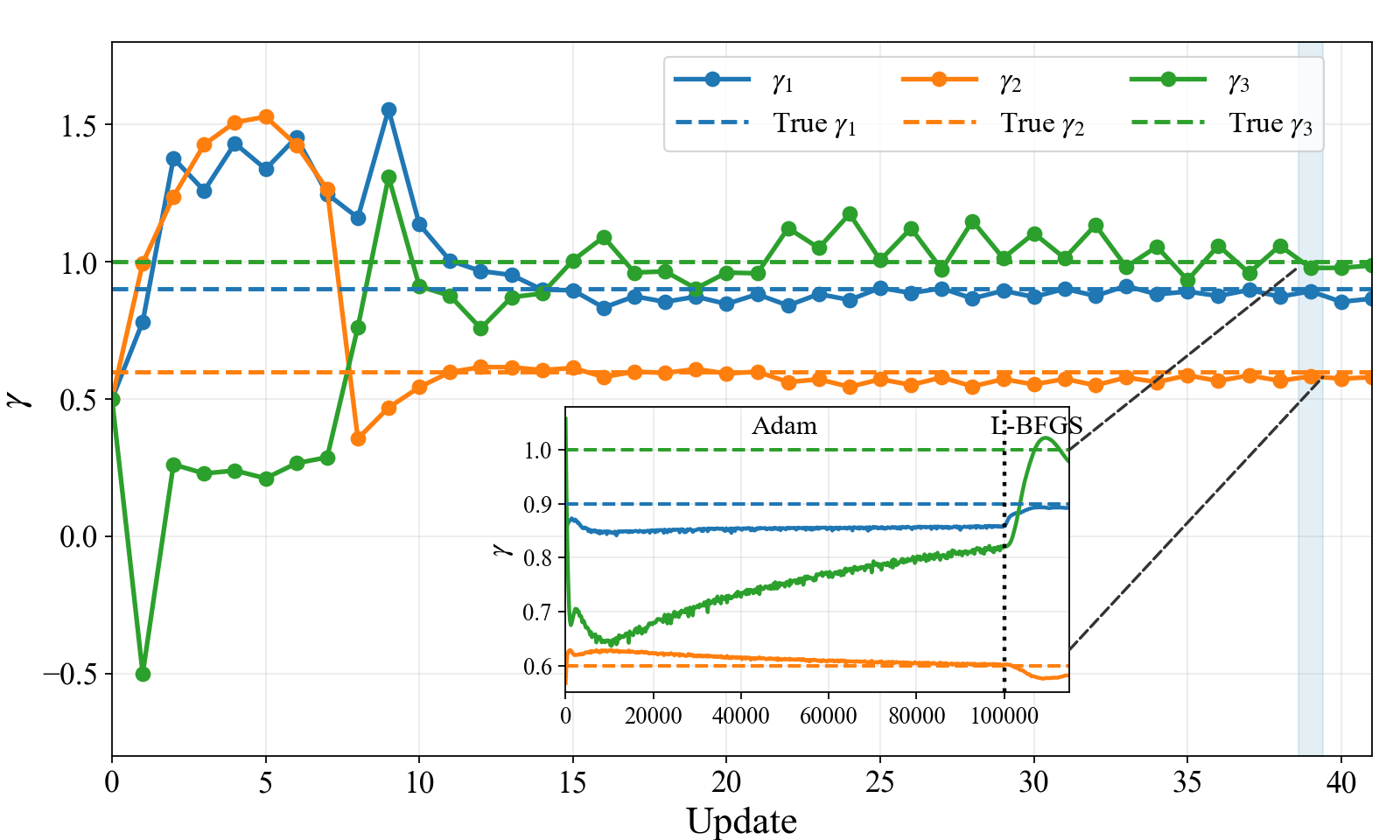}
	\caption{Evolution of the identified upper-story stiffness-reduction
	factors for Case~C5. Horizontal dashed lines denote the ground-truth values,
	while vertical markers indicate successive basis updates. The inset highlights
	the final convergence after switching from Adam to L-BFGS-B.}
	\label{fig:case5_gamma}
\end{figure}

Figure~\ref{fig:case5_gamma} shows the evolution of the identified stiffness-reduction factors during the basis-refinement process. The parameter estimates initially exhibit noticeable fluctuations, reflecting the simultaneous adaptation of the reduced basis, the residual-force estimates, and the structural parameters. As the basis is refined, all three parameters progressively approach their reference values and remain stable during the final optimization stage. These results indicate that the adaptive basis-refinement strategy remains effective in the joint inverse setting.

\begin{figure}[h!]
	\centering
	\subfigure[Brace~8: force--displacement response]{
		\begin{minipage}[b]{0.45\textwidth}
			\includegraphics[width=1\textwidth]{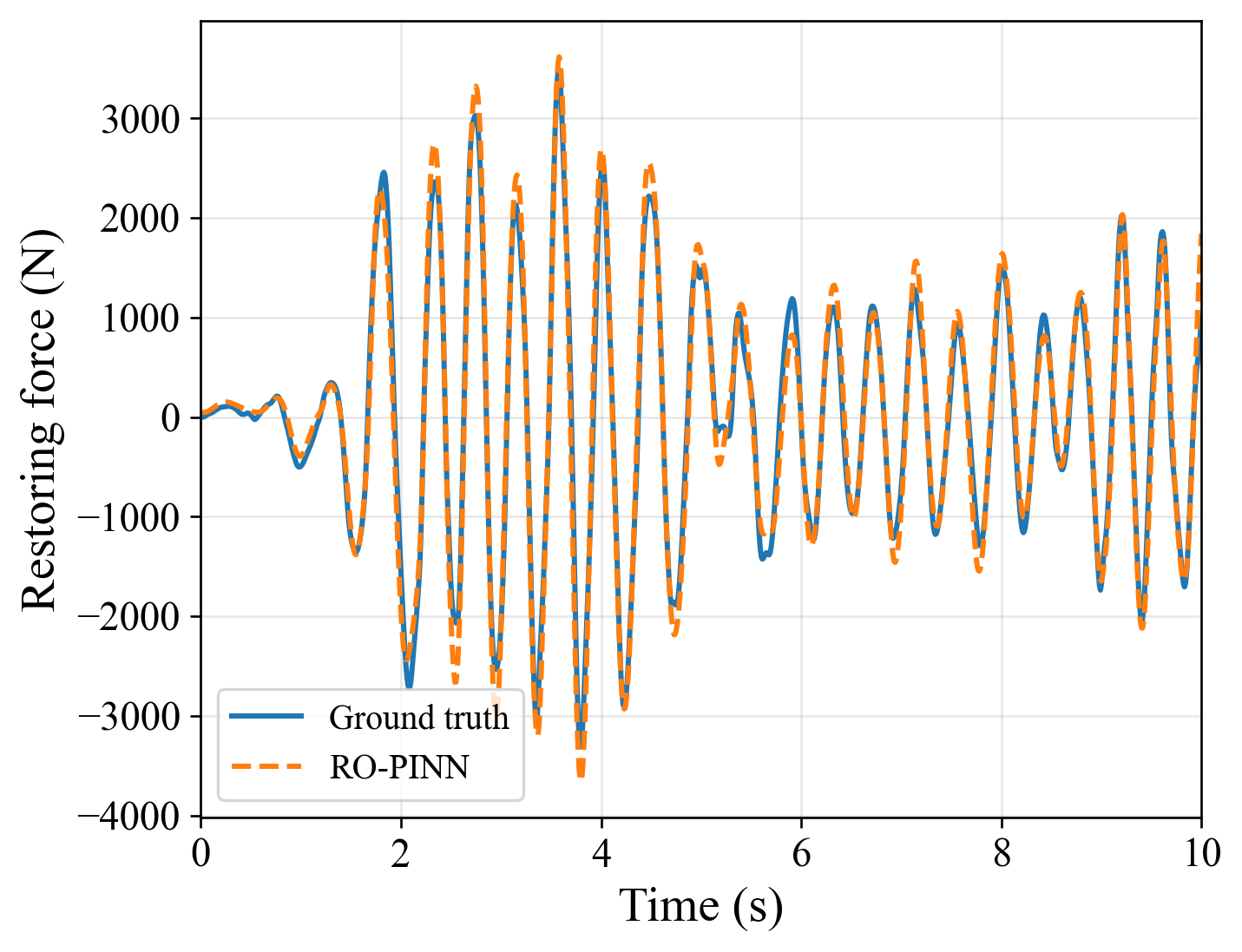}
		\end{minipage}
	}
	\subfigure[Brace~8: force time history]{
		\begin{minipage}[b]{0.45\textwidth}
			\includegraphics[width=1\textwidth]{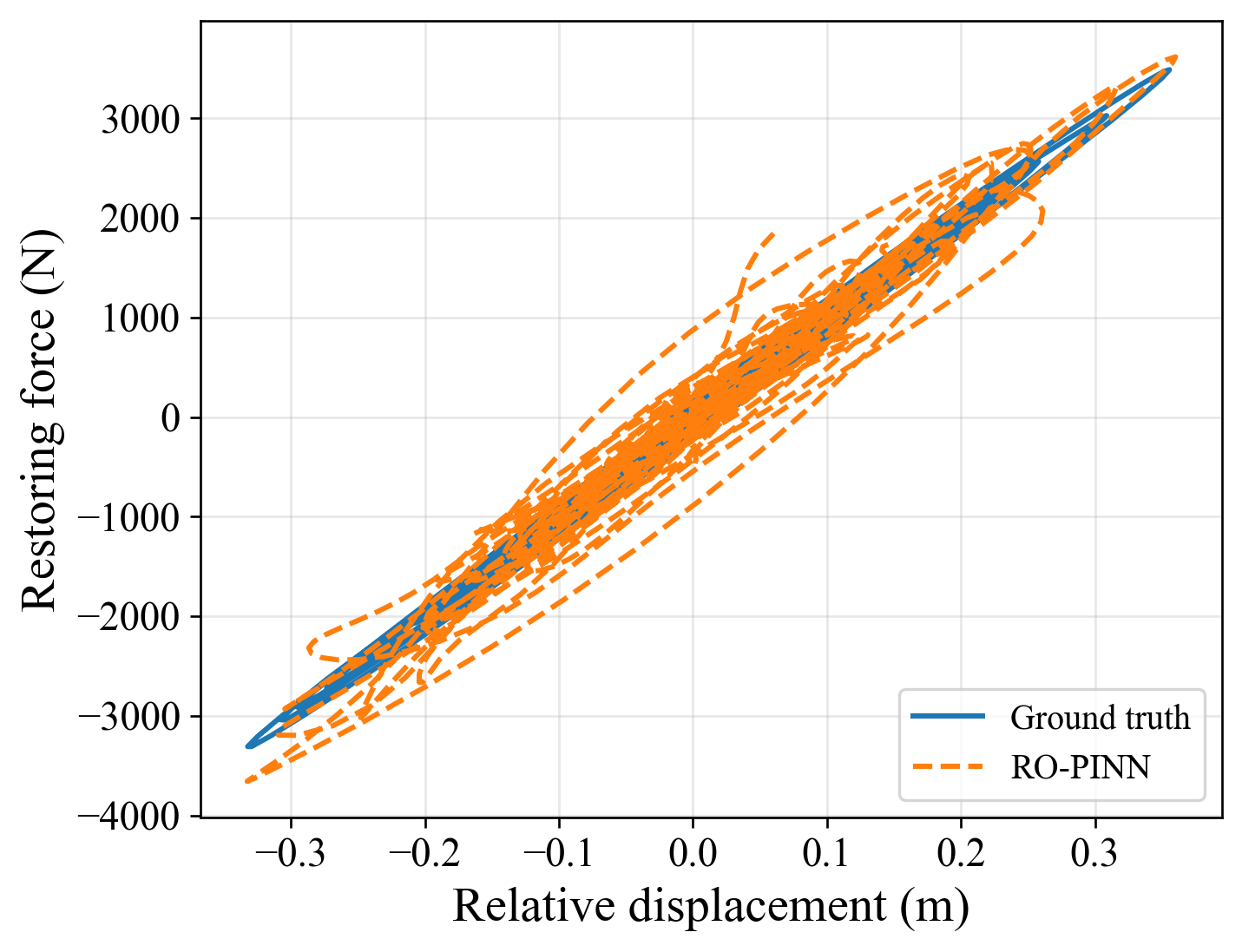}
		\end{minipage}
	}

	\subfigure[Brace~11: force--displacement response]{
		\begin{minipage}[b]{0.45\textwidth}
			\includegraphics[width=1\textwidth]{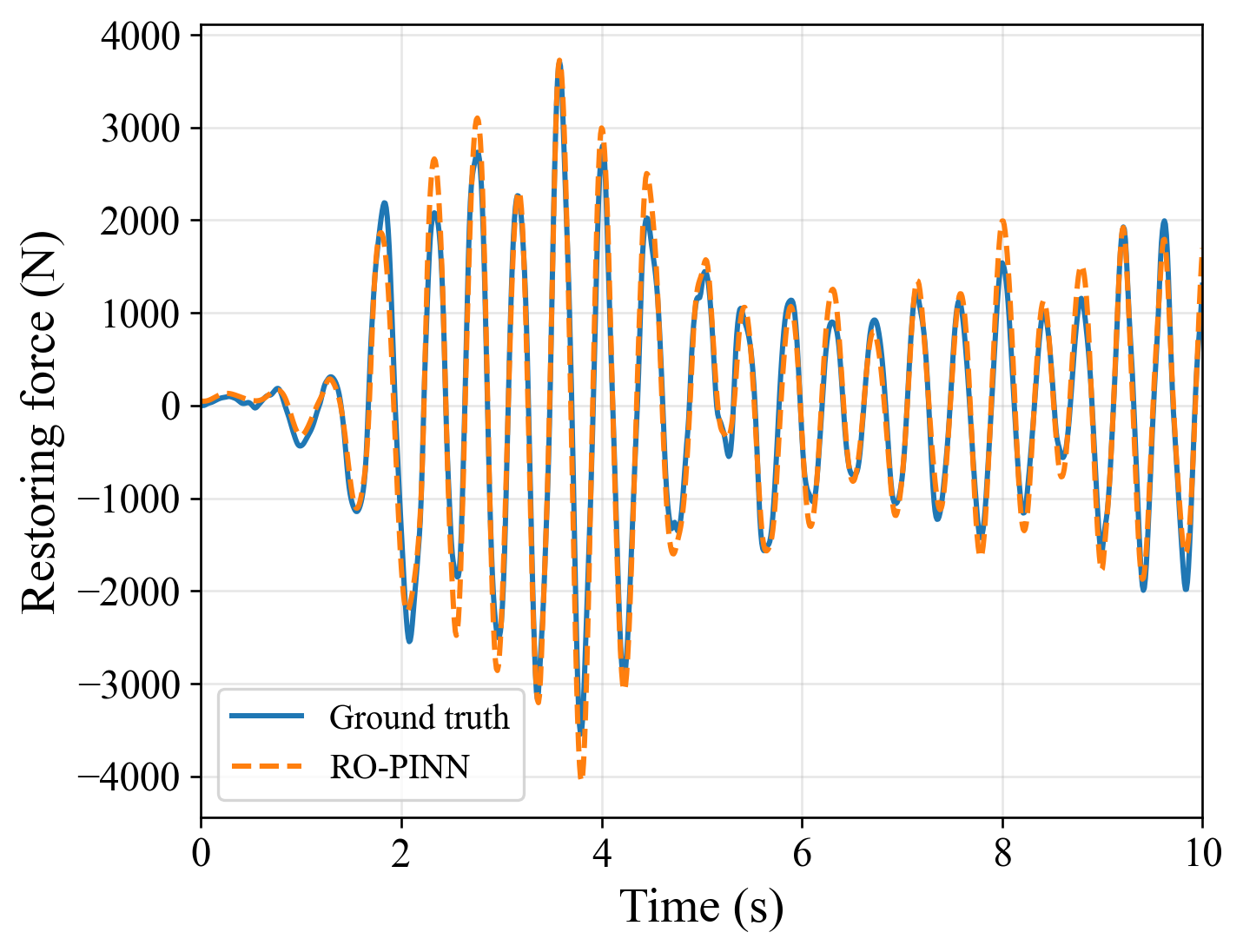}
		\end{minipage}
	}
	\subfigure[Brace~11: force time history]{
		\begin{minipage}[b]{0.45\textwidth}
			\includegraphics[width=1\textwidth]{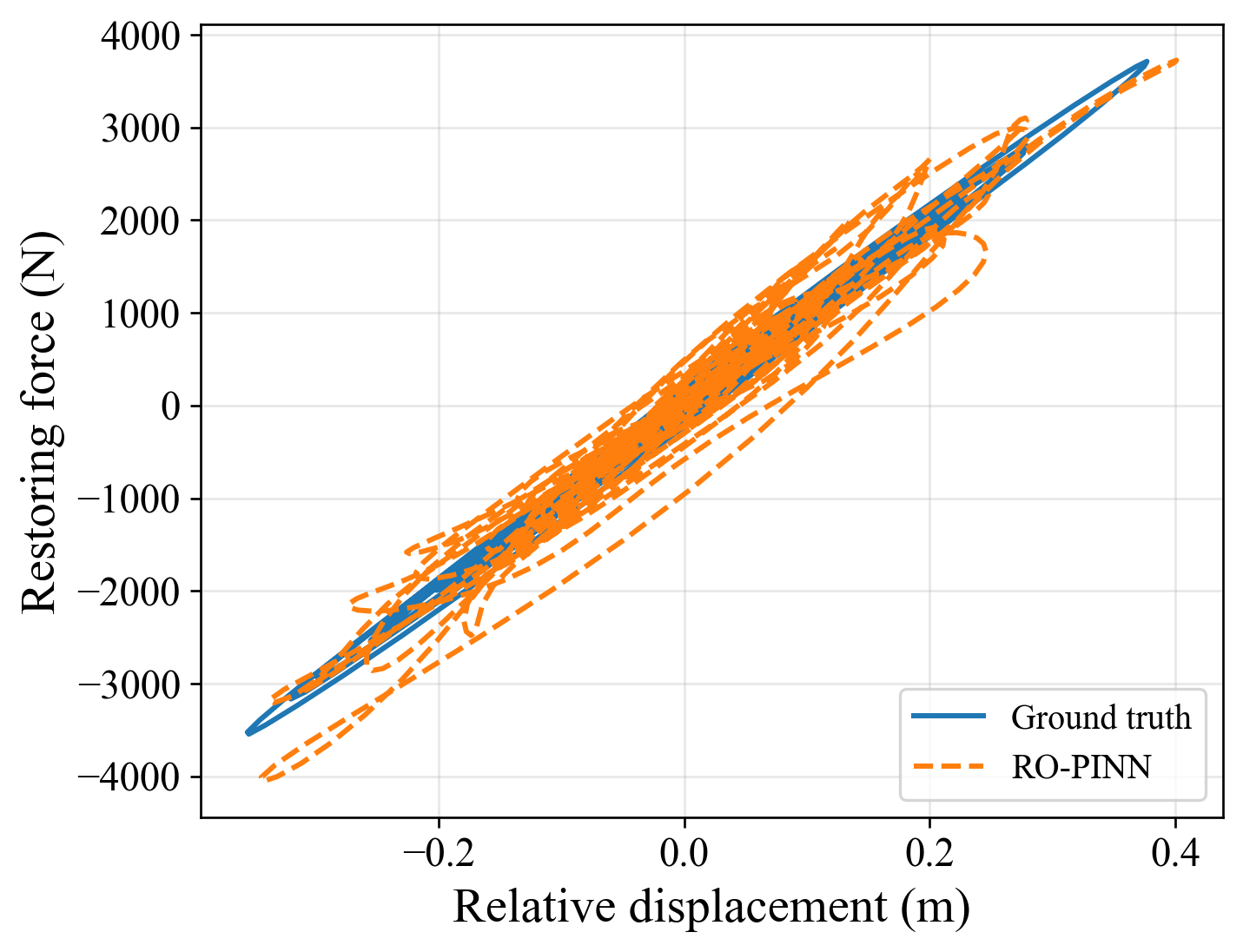}
		\end{minipage}
	}

	\caption{Recovered first-story residual restoring forces for Case~C5:
	comparison between the RO-PINN predictions and the reference solution for two
	representative braces, shown as (a,c) force--displacement hysteresis loops and
	(b,d) force time histories.}
	\label{fig:case5_force}
\end{figure}

Figure~\ref{fig:case5_force} compares the identified first-story restoring forces with the corresponding reference responses for two representative braces. The RO-PINN predictions recover both the force--displacement hysteresis loops and the temporal evolution of the restoring forces with good overall agreement. Minor discrepancies remain in parts of the loading and unloading branches, which is consistent with the greater difficulty of the joint inverse problem relative to the single-unknown cases. Nevertheless, the recovered force histories remain compatible with the simultaneously identified stiffness parameters.

Table~\ref{tab:results_c5} quantifies the contribution of the static-correction enrichment by comparing three reduced bases. The three-mode POD basis is the baseline adopted in the previous cases, whereas the seven-mode POD basis provides a dimension-matched comparison with the proposed three-mode POD basis augmented by four static-correction vectors. Increasing the number of global POD modes improves the results to some extent, but the static-enriched basis consistently provides the most accurate parameter estimates and restoring-force reconstruction. This indicates that the improvement arises not only from a larger reduced dimension, but also from the inclusion of basis vectors tailored to the localized first-story force patterns.

\begin{table}[h!]
\centering
\caption{Performance of Case~C5 using different reduced bases. The comparison isolates the contribution of the proposed static-correction enrichment.}
\label{tab:results_c5}
\footnotesize
\begin{tabular}{lccccc}
\toprule
Basis &
$k_2$ err. (\%) &
$k_3$ err. (\%) &
$k_4$ err. (\%) &
$\mathbf{R}$ err. (\%) &
Time (h) \\
\midrule
POD ($r=3$)
& 45 & 32 & 22 & 24.5 & 7.5 \\

POD ($r=7$)
& 25 & 23 & 12 & 14.5 & 13.0 \\

POD ($r=3$)+4 static
& 1.8 & 2.9 & 3.5 & 7.8 & 11.5 \\
\bottomrule
\end{tabular}
\end{table}

Overall, Case~C5 shows that the proposed RO-PINN framework extends the framework from the separate parameter and residual-force
identification settings to their simultaneous estimation of residual restoring forces and structural parameters. Compared with Cases~C1--C4, the joint problem converges more slowly and yields moderately larger identification errors, reflecting the greater difficulty of inferring multiple sources of uncertainty from the same measurements. At the same time, the static-correction-enriched reduced basis provides a clear improvement for this separated-support joint identification problem.

\subsubsection{Baseline Comparisons: Conventional PINN and Bayesian Model Updating}
\label{s4.4.4}

Two representative baselines are considered for benchmarking: (i) a conventional physics-informed neural network formulated directly in the full-order space (FOM--PINN), and (ii) Bayesian model updating (BMU).

The FOM--PINN directly approximates the full displacement field $\mathbf{u}(t)\in\mathbb{R}^{n}$ and, when required, the associated internal variables, while enforcing the full-order governing equations through automatic differentiation. In the present numerical experiments, this formulation did not converge to sufficiently accurate inverse solutions for the considered cases. The substantially higher output dimension and the more complex optimization landscape of the full-order formulation are consistent with the spectral-bias and training-stiffness difficulties previously reported for PINNs~\cite{wang2021understanding,wang2022and}. The reduced formulation adopted in RO-PINN instead performs learning in a low-dimensional subspace, thereby reducing the number of network outputs and the computational cost of enforcing the governing equations.

BMU is considered as a classical reference for parameter identification under known physics. It combines prior information with likelihood evaluations obtained through repeated forward simulations. For Cases~C1--C2, however, the assumed model intentionally omits the nonlinear brace contribution and contains no parameterization of this missing physics. Consequently, conventional parameter-only BMU cannot recover the corresponding residual restoring force by updating the existing model parameters.

For Cases~C3--C4, BMU is applicable and is implemented using the improved Transitional Markov Chain Monte Carlo (iTMCMC) method~\cite{betz2016transitional}, adapted from the open-source ERA implementation. BMU provides an established reference for parameter identification ~\cite{kamariotis2025consistent}, while its computational cost is dominated by the repeated forward-model evaluations required during sampling. As reported in Table~\ref{tab:results}, both BMU and the adaptively refined RO-PINN recover the unknown stiffness parameters with low errors in Cases~C3--C4. RO-PINN yields mean parameter errors of $1.77\%$ and $3.43\%$, compared with $3.12\%$ and $4.50\%$ for BMU, respectively.

Based on the reported wall-clock times, RO-PINN reaches the final solution in approximately $1.0$ and $0.9$~h for Cases~C3 and C4, whereas the corresponding BMU analyses require $5.8$ and $5.4$~h using four parallel cores. These values indicate a substantial computational advantage for the present examples, although the timing comparison depends on the implementation, parallelization, and forward-model cost.

The two approaches therefore provide complementary capabilities. For parameter identification under a prescribed physical model, both RO-PINN and BMU provide accurate estimates in the considered examples. RO-PINN additionally accommodates residual-force identification when the assumed model is incomplete, as demonstrated in Cases~C1--C2. Case~C5 further combines residual-force and parameter identification within the same inverse problem, whereas the parameter-only BMU formulation considered here does not include an equivalent representation of the missing restoring-force contribution.

\subsection{Challenges and Limitations}

Despite the favorable performance observed in the numerical examples, several limitations remain. First, identifiability becomes more challenging when different sources of uncertainty, such as frame stiffness loss, brace degradation, and unmodeled nonlinear effects, produce similar dynamic signatures. This issue is particularly relevant for localized changes, whose influence may be weak in the measured global response, increasing ambiguity between physical changes, correlated parameters, and reduced-basis mismatch.

Second, the amount, type, and spatial distribution of measurements strongly affect reconstruction and identification accuracy. Displacement measurements mainly reflect lower-frequency structural response, whereas acceleration measurements are generally more sensitive to measurement noise. In this study, a combination of displacement and acceleration measurements is used to provide complementary information under sparse sensing. Additional sensors or other response quantities, such as strain measurements, may further improve observability and identification accuracy.

Finally, joint residual-force and parameter identification remains more challenging because the two classes of unknowns may produce similar effects in the measured response and therefore partially compensate for one another during the inverse solution. Case~C5 provides a first demonstration in which the residual-force contribution and the unknown stiffness parameters act on different structural regions. The static-correction enrichment improves the representation of the localized residual-force effects, although the joint problem converges more slowly and yields larger identification errors than the corresponding single-setting cases. Extending the framework to the more challenging same-support configuration, where residual forces and uncertain parameters coexist within the same structural region, remains an important direction for future investigation.

%%%%%%%%%%%%%%%%%%%%%%%%%%%%%%%%%%%%%%%%%%%%%%%%%%%%%%%%%%%%%%%%%%%%%%
\section{Conclusion}
\label{s5}

This study presented RO-PINN, a Reduced-Order Physics-Informed Neural Network with adaptive basis refinement for inverse structural problems under known and incomplete physics. By combining physics-informed neural networks with projection-based reduced-order modeling, the proposed framework performs inverse identification in a low-dimensional representation while allowing the reduced basis to adapt as the inferred physical quantities evolve. The formulation accommodates parameter identification, residual-force identification, and their simultaneous estimation, and can additionally include reduced internal variables governed by known evolution laws.

The framework was demonstrated on a four-story steel frame with nonlinear hysteretic braces through five numerical cases. For residual-force identification under incomplete physics, RO-PINN recovered the missing nonlinear restoring-force contribution while reconstructing structural responses at both measured and unmeasured degrees of freedom. For parameter identification under known physics, adaptive basis refinement substantially reduced stiffness-parameter errors and maintained accurate response reconstruction under sparse and noisy measurements. The numerical results further showed that updating the reduced basis as the inverse estimates evolve is important for limiting basis-mismatch effects and maintaining consistency between the reduced model and the identified system.

The joint-identification case further demonstrated simultaneous recovery of localized residual restoring forces and upper-story stiffness parameters when the two classes of unknowns act on different structural regions. Static-correction enrichment improved the representation of the localized force effects and yielded more accurate identification than dimension-matched global POD enrichment alone. This result extends the framework beyond separate parameter and residual-force identification and provides a first demonstration of joint residual-force and parameter identification within the proposed reduced-order physics-informed formulation.

Several challenges remain. Identifiability may deteriorate when different uncertainty sources produce similar dynamic signatures, particularly under localized changes and sparse sensing. Extending the joint formulation to the more challenging same-support setting, where residual forces and uncertain parameters coexist within the same structural region and may compensate for one another, is therefore an important direction for future work. Further extensions include uncertainty quantification for the identified quantities and
reconstructed responses, experimental validation, and computational strategies for scaling the framework to larger structural systems and real-time structural health monitoring and digital-twin applications.

\section*{CRediT authorship contribution statement}

\textbf{Rui Zhang}: Conceptualization, Methodology, Software, Formal analysis, Writing - original draft, Writing - review and editing. \textbf{Konstantinos Vlachas}: Conceptualization, Methodology, Formal analysis, Writing - review and editing. \textbf{Eleni Chatzi}: Conceptualization, Writing - review and editing, Supervision.

\section*{Declaration of Competing Interest}

The authors declare that they have no known competing financial interests or personal relationships that could have appeared to influence the work reported in this paper.

\section*{Acknowledgments}

This work was supported by an ETH Zurich Postdoctoral Fellowship Grant (No. 24-1 FEL-016, Host Project), titled “Uncertainty Quantification–Integrated Scientific Machine Learning (UQ-SciML) Framework for Structural Identification.” Furthermore, the authors \textbf{KV, EC} gratefully acknowledge the funding from the European Commission under the Horizon Europe funding guarantee, for the project ‘TURING - Trustworthy Unified Robust Intelligent Generative Systems’ (grant agreement No. 101215032).

\section*{Data availability}

Data will be made available on request. 




\bibliographystyle{elsarticle-num} 
\bibliography{mybibfile}

%\end{linenumbers}
\end{document}